\documentclass[prb, reprint, 9pt, superscriptaddress,notitlepage, nofootinbib, longbibliography,
floatfix]{revtex4-2}

\usepackage{natbib}
\usepackage{graphicx}
\usepackage{epsfig}
\usepackage{amsfonts} 
\usepackage{amsmath} 
\usepackage{amssymb}
\usepackage{diagbox}
\usepackage{dsfont}
\usepackage{bm}
\usepackage{braket}
\usepackage{color}
\usepackage{physics} 
\usepackage{bbm}
\usepackage{hyperref}
\usepackage{subcaption}
\usepackage{comment}
\usepackage[export]{adjustbox}  
\usepackage{enumitem}

\renewcommand{\raggedright}{\leftskip=0pt \rightskip=0pt plus 0cm}
\newcommand{\be}{\begin{equation}}
	\newcommand{\ee}{\end{equation}}
\newcommand{\ba}{\begin{eqnarray}}
	\newcommand{\ea}{\end{eqnarray}}

\newcommand{\mA}{\mathcal{A}}
\newcommand{\mB}{\mathcal{B}}

\definecolor{LinkColor}{rgb}{0,0,1}
\hypersetup{
	colorlinks=true,
	citecolor=LinkColor,
	linkcolor=LinkColor,
	urlcolor=LinkColor
}

\definecolor{gr}{rgb}{0,0,0}

\begin{document}
	
 \title{Disentangling transitions in topological order induced by boundary decoherence}

\author{Tsung-Cheng Lu}

\affiliation{Perimeter Institute for Theoretical Physics, Waterloo, Ontario N2L 2Y5, Canada}
\affiliation{Department of Physics, Boston University, Boston, MA, 02215, USA}

\begin{abstract}
We study the entanglement structure of topological orders subject to decoherence on the bipartition boundary. Focusing on the toric codes in $d$ space dimensions for $d=2,3,4$, we explore whether the boundary decoherence may be able to induce a disentangling transition, characterized by the destruction of mixed-state long-range entanglement across the bipartition, measured by topological entanglement negativity. A key insight of our approach is the connection between the negativity spectrum of the decohered mixed states and emergent symmetry-protected topological orders under certain symmetry-preserving perturbation localized on the bipartition boundary. This insight allows us to analytically derive the exact results of entanglement negativity without using a replica trick. 
 \end{abstract}
	
	\maketitle

	
	{
		\hypersetup{linkcolor=black}
	}

Quantum many-body systems with topological orders \cite{wen2004quantum}, owing to their long-range entanglement structure, may be used as topological quantum error-correcting codes \cite{kitaev2003fault,dennis2002}, where the encoded logical information can be protected and processed reliably under sufficiently weak environmental noise. Correspondingly, there exists a decodability transition at a certain critical noise threshold, above which the information is corrupted and no longer decodable. A seminal example is the 2d toric code, where the decodability transition can be understood from the order-disorder transition of the 2d random bond Ising model \cite{dennis2002}.

While the decodability transition has long been a familiar concept in the field of quantum information, it possesses several intriguing features from the standpoint of quantum phases of matter. For one thing, this transition cannot be detected via any linear observables $O$ of the noisy mixed-state density matrix $\rho$ \cite{fan_2023_toric}. Namely, the observable of the form $\tr( \rho O)$ is a smooth function of the noise rate. This observation therefore motivates several outstanding questions. Does the decodability transition separate two distinct mixed-state quantum phases? How should one classify phases of matter in mixed states? Perhaps most intriguingly, does the decodability transition correspond to a certain intrinsically `quantum' phase transition, which only manifests in the entanglement structure, as opposed to any physical observables, of the noisy mixed state? These questions have provide new opportunities to explore mixed-state phases and their transitions \cite{spt_Schuch_2022,wang_spt_2023,lee2022_spt,criticality_2023_xu,fan_2023_toric,bao2023mixed,Lu_mixed_feedback_2023,hsieh_2023_criticality,grover_2023_separability_toric,chen2023symmetry,su2023higher,sang2023mixed,wang2023intrinsic,ma_2023topological_mixed,hsin2023anomalies,xu_chern_2024,xu_2024_anyon_decoherence,mong_2024_replica,jian_2024_duality,lyons2024understanding,lee2024exact,chen2024unconventional,Zhen_2024_spt,Turzillo_2024_spt,Yimu_2024_tn,sohal2024noisy,wang2024anomaly}.

\begin{figure}
\centering
\begin{subfigure}{0.30\textwidth}
\includegraphics[width=\textwidth]{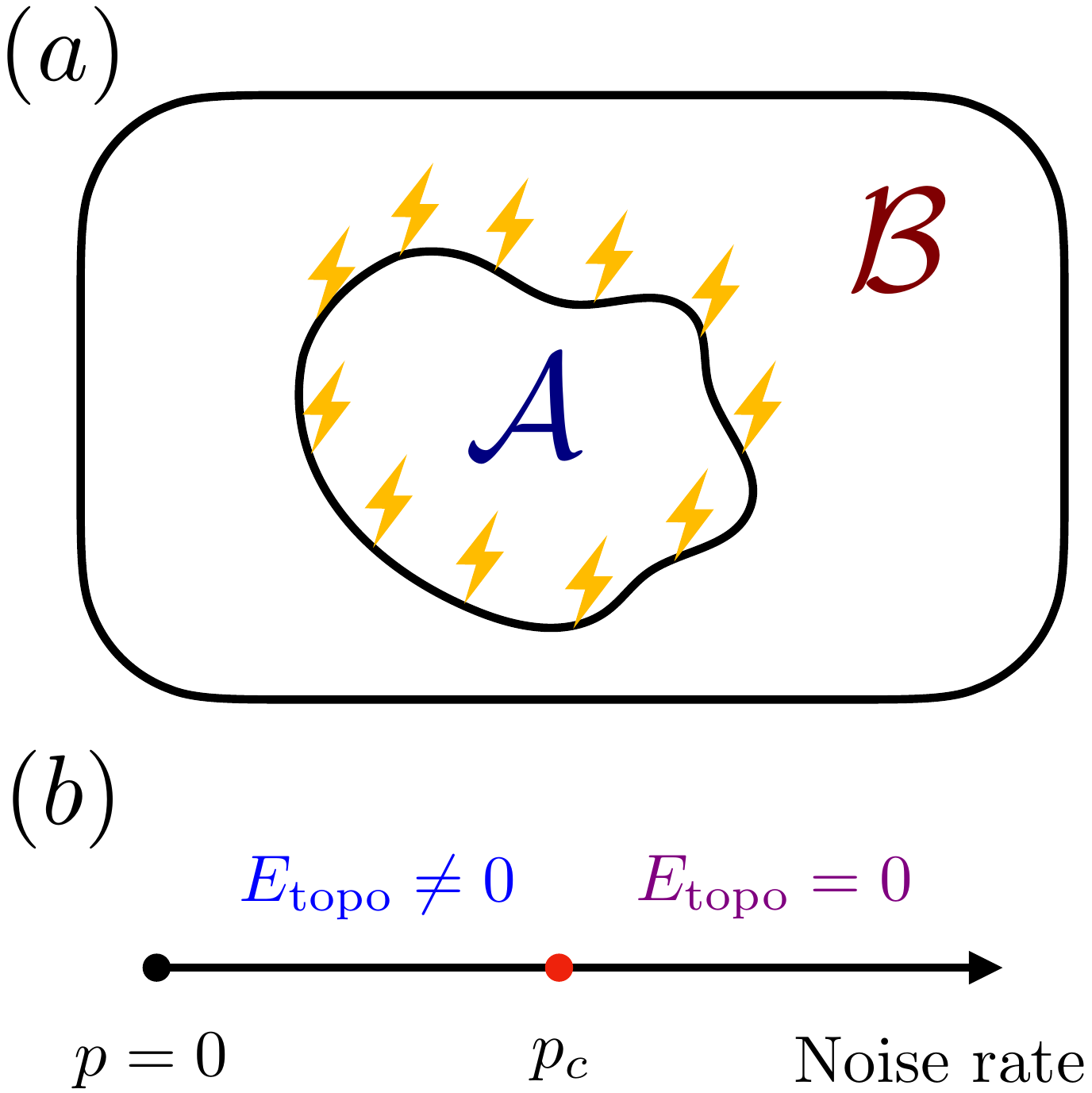}
\end{subfigure}
\caption{(a) A topologically-ordered state subject to the decoherence acting \textit{only} on the bipartition boundary between $\mA$ and $\mB$. (b) Tuning the noise rate $p$ may induce a disentangling transition at a critical error rate $p_c$, above (below) which the long-range entanglement between $\mA$ and $\mB$, measured by topological entanglement negativity $E_{\text{topo}}$, is zero (non-zero).}
\label{fig:main_fig}
\end{figure}

In this work, we consider a setup, where the entanglement structure of certain decohered many-body states can be completely understood. We consider a topologically-ordered system divided into two regions $\mA$ and $\mB$, and the decoherence \textit{only} acts on the bipartition boundary (see Fig.\ref{fig:main_fig}). In this case, one expects the entanglement structure and topological order in the respective bulks of $\mA$ and $\mB$ to remain intact for arbitrary decoherence strength. Therefore, the essential question is how the boundary decoherence affects the entanglement structure across the interface. In particular, does there exist a `disentangling transition’ at a certain critical decoherence strength, across which the long-range entanglement between $\mA$ and $\mB$ is destroyed?  We will address this question for the toric code models in various space dimensions and explore the entanglement structure via entanglement negativity \cite{peres1996,horodecki1996,eisert99,vidal2002}, an entanglement measure for mixed states. Specifically, we will employ topological entanglement negativity, a subleading contribution of entanglement negativity, to diagnose long-range entanglement and characterize the disentangling transition.

Our analytical approach is built on Ref.\cite{lu2022_lre}, which points out a correspondence between certain bulk topological orders and symmetry-protected topological (SPT) orders \cite{spt_1d_2011,spt_2011} emerging from the calculation of entanglement negativity. Recall that entanglement negativity of a bipartite mixed state $\rho$ is defined as $E_N= \log \abs{\rho^\Gamma}_1 =  \log \left(\sum_i\abs{\lambda_i}  \right)$, where $\rho^\Gamma$ results from taking a partial transpose (i.e. the transpose only acting on the subregion $\mA$) on the density matrix $\rho$, and $\{\lambda_i \}$ is the eigenspectrum  $\rho^\Gamma$, also dubbed negativity spectrum. It was found that the negativity spectrum of the ground-subspace density matrix of topological codes in $d$ space dimensions exactly corresponds to an SPT wave function localized on the $d-1$ dimensional bipartition boundary. Taking the 2d ($\mathbb{Z}_2$) toric code as an example, the negativity spectrum corresponds to the wave functions in symmetry-charge basis (or equivalently, the strange correlators) of a 1d cluster state that exhibits a  $\mathbb{Z}_2\times \mathbb{Z}_2$ SPT order \cite{Raussendorf_2001_ghz}. Heuristically, this follows from the fact that taking a partial transpose on the toric-code ground-space density matrix proliferates $e$-particles and $m$-particles along the bipartition boundary with non-trivial mutual braiding statistics, and this is precisely captured by the braiding phase between two types of the $\mathbb{Z}_2$ symmetry charges encoded in the SPT.

Built on this result, we find that introducing boundary decoherence amounts to introducing certain symmetry-preserving perturbations on the SPT, and the wave functions of this perturbed SPT completely characterize the negativity spectrum of the decohered mixed state. Importantly, this allows us to analytically compute the entanglement negativity and extract the topological contribution (i.e. topological entanglement negativity) without resorting to a replica trick; for the topological order in $d$ dimensions subject to the boundary decoherence, the entanglement negativity relates to the free energy difference associated with annihilating domain walls in \textit{translationally invariant} statistical mechanics models in $(d-1)$-dimensions. While our analytical approaches apply to any qubit-stabilizer models, we will focus on the discussion of the toric codes in various space dimensions, which we summarize below. 

For the 2d toric code subject to the boundary decoherence, the entanglement negativity relates to the free energy of the 1d Ising model. Due to the absence of a finite-temperature transition in the 1d Ising model, topological entanglement negativity $E_{\text{topo}}$ of the decohered toric code remains the non-zero quantized value $\log 2$ for any non-maximal noise rate. Namely,  $E_{\text{topo}}$ becomes zero identically only at the maximal decoherence strength $p=\frac{1}{2}$. 

For the 3d toric code, when introducing the boundary noise that creates loop-like defects, the emergent statistical mechanics model is a $Z_2$ gauge theory in 2d, which does not exhibit any finite-temperature transition. Correspondingly, $E_{\text{topo}}$ remains $\log 2$ for any non-maximal noise strength. On the other hand, under the decoherence that creates point-like defects, the entanglement structure is mapped to a 2d Ising model, whose finite-temperature transition indicates the disentangling transition of the decohered toric code at a certain critical decoherence strength $p_c$, below which $E_{\text{topo}} = \log 2$, and above which, $E_{\text{topo}} = 0$. 

Finally, for the 4d toric code subject to the boundary decoherence, the emergent statistical mechanics model is a $Z_2$ gauge theory in 3d, which exhibits a confinement-deconfinement transition at finite temperature. Correspondingly, there exists a disentangling transition at a certain critical decoherence strength $p_c$, below which $E_{\text{topo}} = 2\log 2$, and above which, $E_{\text{topo}} = 0$. 

\textbf{\textit{Topological entanglement negativity}} - In this work, we will employ topological entanglement negativity to characterize the bipartite long-range entanglement of topological orders subject to boundary decoherence. Consider a smooth bipartition boundary of size $L$, one expects the bipartite mixed-state entanglement measured by entanglement negativity follows the form: $E_N= E_{\text{local}}-  E_{\text{topo}}$. $E_{\text{local}}$ captures the short-range entanglement along the bipartition boundary and exhibits an area-law scaling $E_{\text{local}}  \sim  \alpha_{d-1 } L^{d-1}$ to leading order in $L$, and $E_{\text{topo}}$ is the \emph{topological entanglement negativity}, a universal constant contribution proposed to diagnose the long-range entanglement. Such a scaling form is first proposed in Ref.\cite{lu2019structure} based on the locality structure of many-body systems, and later demonstrated for finite-temperature Gibbs states \cite{Lu_topo_nega_2020,lu2022_lre} and certain topological order under decoherence \cite{fan_2023_toric,wang2023intrinsic,lyons2024understanding}. When the system of interest is described by a pure state, the negativity coincides with Renyi-1/2 entanglement entropy \cite{vidal2002}, and the topological entanglement negativity reduces to topological entanglement (Renyi) entropy \cite{Kitaev06_1}.

\textbf{\textit{2d toric code}} - Here we discuss the negativity spectrum and the exact result of entanglement negativity in the 2d toric code subject to Pauli-Z noise on the bipartition boundary \footnote{We only present the result for Pauli-Z noise as Pauli-X noise will lead to the same mixed-state entanglement structure due to the duality between $A_s$ and $B_p$ stabilizers.}. We will discuss the essential physics and outline the derivation in the main text. A unified, detailed calculation for toric codes of various space dimensions is presented in Appendix.\ref{append:spectrum}. 

The toric-code Hamiltonian reads $H= -\sum_s A_s  - \sum_p B_p$, where the star stabilizer $A_s$ is a product of Pauli-Zs on four edges emanating from a site $s$, and plaquette stabilizer $B_p$ is a product of Pauli-Xs on four edges around a plaquette $p$. The density matrix that describes the ground subspace of the toric code is $\rho_0  \propto \prod_s \frac{1+A_s}{2}  \prod_p \frac{1+ B_p}{2}$.

Consider a 1d closed bipartition boundary of size $L$ \footnote{The boundary can either be contractible (e.g. Fig.\ref{fig:main_fig}a), or non-contractible, e.g. winding around $x$ direction of the 2d lattice with periodic (open) boundary condition in $x$ $(y)$ direction (the topology of a cylinder). Our calculation applies to both cases.}, there are $2L$ alternating star and plaquette stabilizers along  the boundary, with the dashed line denoting the bipartition boundary: 

 \begin{equation*}
 \includegraphics[width=6cm]{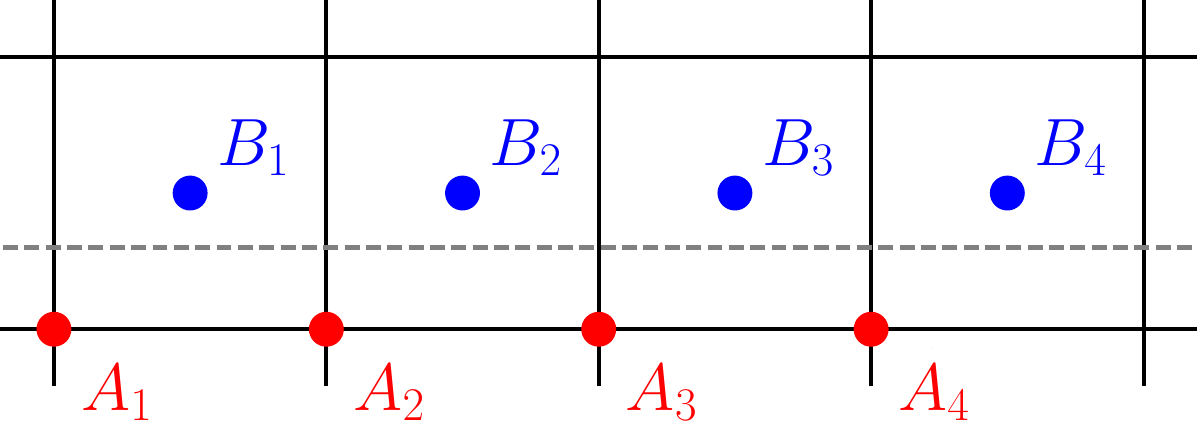}
 \end{equation*}

As will be convenient, we factorize the mixed state into a bulk part and a boundary part: $\rho_0 \propto \rho_{\text{bulk}} \rho_{\partial,0}$, where the boundary part $\rho_{\partial,0} = \prod_{i=1}^L \frac{1+ A_i}{2} \prod_{i=1}^L \frac{1+ B_i}{2}$ involves the boundary stabilizers, and $\rho_{\text{bulk}} = \prod_{s\in \text{bulk}}\frac{1+ A_s}{2} \prod_{p \in \text{bulk}} \frac{1+ B_p }{2}$ involves the bulk stabilizers.

Now we consider the decoherence that only creates the boundary $B_i$ excitations by applying Pauli-Z operators on edges shared by two adjacent $B_i$s with probability $p_z$. This leads to a decohered state $\rho= \prod_i N_{z,i} [\rho_0]$, where $N_{z,i}[~]$ is a local noise channel with the action $N_{z,i}[\rho] = (1-p_z) \rho + p_z Z_i \rho  Z_i$, and $Z_i$ denotes the Pauli-Z operator on the edge shared between $B_i$ and $B_{i+1}$. To see the effect of the noise channel, we expand the initial boundary density matrix as a sum over all possible stabilizer strings: $\rho_{\partial,0}  \propto \sum_{ a,b} \prod_i A_i^{a_i}\prod_i B_i^{b_i}$ with $a\equiv \{ a_i\}$, $b\equiv \{ b_i\}$, and each $a_i, b_i\in \{ 0,1\}$ denote the absence/presence of $A_i, B_i$. 

Applying the noise channel amounts to introducing a `tension' so that the weight of a given stabilizer string $\prod_iA_i^{a_i}\prod_i B_i^{b_i }$ is suppressed exponentially in the number of Pauli-Xs acting on the qubits that are subject to decoherence, denoted as $W(X$). Namely,  $\rho_{\partial}  \propto  \sum_{ a,b  } \prod_iA_i^{a_i}\prod_i B_i^{b_i } (1-2p_z)^{W(X)}$, and $W(X)$ can be computed as follows. For a given edge, the Pauli-X exists when only one of the two neighboring plaquette stabilizers $B_i, B_{i+1}$ is present. Therefore, $W(X)  =  \frac{1}{2} \sum_i( 1-  \tau_i \tau_{i+1})$, with $\tau_i = 1-2b_i \in \{ 1, -1   \}$. By defining $\beta_z  = - \frac{1}{2}\log (1-2p_z)$, the boundary decohered state can be written as 

\begin{equation}
\rho_{\partial} \propto   \sum_{a,b} \prod_i  A_i^{a_i} \prod_i B_i^{b_i} e^{\beta_z  \sum_{i=1}^{L}\tau_i \tau_{i+1} }, 
\end{equation}
where the Ising interaction among $\{b_i\}$ (or equivalently, among $\{\tau_i\}$) incorporates the effect of the decoherence.

To proceed, one can expand the decohered mixed state $\rho \propto  \rho_{\text{bulk}} ~\rho_{\partial}$ into a sum over all possible stabilizer strings $\prod_{s} A_s^{a_s} \prod_{p} B_p^{b_p}$, with $a_s, b_p  =0, 1$. Then taking a partial transpose assigns a sign ($1 \text{ or } -1$) depending on whether the number of Pauli-Y operators in the subregion $\mA$ is even or odd. This number is only determined by the choice of the boundary stabilizers, and therefore, taking a partial transpose on a subregion acts non-trivially only on the boundary density matrix, i.e. $\rho^{\Gamma}  \propto \rho_{\text{bulk} } \rho_{\partial}^{\Gamma}$. Specifically, taking a partial transpose on the boundary stabilizer string gives $\left(\prod_i A_i^{a_i}\prod_i B_i^{b_i}  \right)^{\Gamma}  = \psi (a,b) \prod_i A_i^{a_i}\prod_i B_i^{b_i}$ with the sign
$\psi(a,b) = \prod_i (-1)^{a_ib_i+ b_i a_{i+1}  }   = 1 \text{ or }-1$, because the simultaneous presence of two neighboring star and plaquette boundary stabilizers contributes to a $-1$ associated with a single Pauli-Y in the subregion $\mA$ \footnote{Alternatively, the $-1$ sign follows from the fact that two neighboring star and plaquette boundary stabilizers anticommute when restricted in the subregion $\mA$.}.

As a result, one finds $\rho^{\Gamma}  \propto\rho_{\text{bulk} } \rho_{\partial}^{\Gamma}$, with 
\begin{equation}
\rho_{\partial}^\Gamma \propto   \sum_{a,b} \psi(a,b) \prod_i A_i^{a_i }\prod_i B_i^{b_i} e^{\beta_z  \sum_{i=1}^{L} \tau_i\tau_{i+1} }
\end{equation}

Since the partially-transposed matrix $\rho^{\Gamma}$ involves only stabilizers, all the eigenvalues (i.e. negativity spectrum of the decohered mixed state) can be obtained by substituting the stabilizers to be 1 or -1. To have non-vanishing eigenvalues, $A_s$ and $B_p$ in the bulk are fixed to be 1. On the other hand, the boundary stabilizers $A_i$ and $B_i$ can be 1 or -1, thereby determining the negativity spectrum. In the following, we will abuse the notation by treating $\rho^{\Gamma}$ as the spectrum (determined by all possible choices of the boundary stabilizer values), rather than a matrix.

It is illuminating to understand the negativity spectrum using a dual description in terms of the variables $\{a_i\}$ and $\{b_i\}$. Introducing a fictitious Hilbert space of $2L$ qubits along the 1d bipartition boundary, with the Z-basis state $\ket{a,b} \equiv  \ket{a_1, b_1, a_2, b_2,\cdots}$ where $Z_{A,i}\ket{a,b}= (-1)^{a_i}\ket{a,b}$ and $Z_{B,i}\ket{a,b}= (-1)^{b_i}\ket{a,b}$, one can define a state $\ket{\psi}  \propto  \sum_{a,b} \psi(a,b)  \ket{a,b} $, and a trivial state $\ket{+} \propto \sum_{a,b}\ket{a,b}$. Then the negativity spectrum can be expressed as

\begin{equation}\label{eq:2d_spectrum}
\rho^{\Gamma} \propto  \bra{+} \prod_i Z_{A,i}^{\frac{1-A_i}{2}}  \prod_i Z_{B,i}^{\frac{1-B_i}{2}}   e^{\beta_z \sum_{i} Z_{B,i}Z_{B,i+1}}\ket{\psi},
\end{equation}
where distinct eigenvalues in the negativity spectrum are given by distinct choices of $A \equiv \{A_i \}$, $B \equiv \{ B_i\}$, which in turn determines the patterns of inserting $Z_{A,i}$ and $Z_{B,i}$, sandwiched in between $\ket{+}$ and $e^{\beta_z \sum_{i} Z_{B,i}Z_{B,i+1}}\ket{\psi}$ in the fictitious Hilbert space. $\ket{\psi}$ is the 1d cluster state with the parent Hamiltonian $H= - \sum_{i=1}^L  Z_{A,i}X_{B,i}Z_{A,i+1} -\sum_{i=1}^LZ_{B,i}X_{A,i+1}Z_{B,i+1}$, which is known to exhibit an SPT order protected by a $\mathbb{Z}_2\times \mathbb{Z}_2$ symmetry generated by $\prod_{i=1}^L X_{A,i}$ and $\prod_{i=1}^L X_{B,i}$. 

As such, the negativity spectrum can be understood as the wave functions of the state $e^{\beta_z \sum_{i} Z_{B,i}Z_{B,i+1}}\ket{\psi}$ in the X-basis, namely $\prod_i Z_{A,i}^{\frac{1-A_i}{2}}  \prod_i Z_{B,i}^{\frac{1-B_i}{2}} \ket{+}$. Alternatively, it can be understood as the strange correlators \cite{You_strange_correlator_2014}, i.e. the symmetry-charge operators $ \prod_i Z_{A,i}^{\frac{1-A_i}{2}}  \prod_i Z_{B,i}^{\frac{1-B_i}{2}}$ inserted in between $\ket{+}$ and $e^{\beta_z \sum_{i} Z_{B,i}Z_{B,i+1}}\ket{\psi}$.

Using the condition $Z_{B,i}X_{A,i+1}Z_{B,i+1} \ket{\psi } = \ket{\psi}$, the negativity spectrum can be simplified as 
\begin{equation}\
\begin{split}
\rho^{\Gamma} &\propto  \bra{+} \prod_i Z_{A,i}^{\frac{1-A_i}{2}}  \prod_i Z_{B,i}^{\frac{1-B_i}{2}}   e^{\beta_z \sum_{i}X_{A,i}}\ket{\psi}\\
&= \bra{+} \prod_i Z_{A,i}^{\frac{1-A_i}{2}}  \prod_i Z_{B,i}^{\frac{1-B_i}{2}}   \ket{\psi} e^{\beta_z \sum_{i}A_i}.
\end{split}
\end{equation} 
Finally, defining $\chi(A,B) = \frac{  \bra{+} \prod_i Z_{A,i}^{\frac{1-A_i}{2}}  \prod_i Z_{B,i}^{\frac{1-B_i}{2}}   \ket{\psi}}{\bra{+}   \ket{\psi}  }$, the negativity spectrum reads

\begin{equation}\label{main:2dspectrum}
\rho^{\Gamma} = \chi(A,B) \frac{e^{\beta_z \sum_i A_i}  }{Z}. 
\end{equation}
where $Z$ is the normalization constant to ensure the negativity spectrum sums to 1 (since the trace of a matrix is invariant under partial transpose). $\chi(A,B)$ is non-zero only when the configurations $A, B$ satisfy $\prod_i A_i= \prod_i B_i = 1$ due to the $\mathbb{Z}_2 \times \mathbb{Z}_2$ symmetry of $\ket{\psi}$, and the non-vanishing $\chi(A,B)$ takes the value $\pm 1 $ depending on whether there exists non-trivial braiding between $A$ and $B$. For instance, consider the configuration with only $A_1,  B_2, A_3,  B_4$ equal $-1$, which corresponds to the following diagram by connecting two points with $A_i=-1$ via a red string and two points with $B_i=-1$ via a blue string: 
\begin{equation*}
\includegraphics[width=6cm]{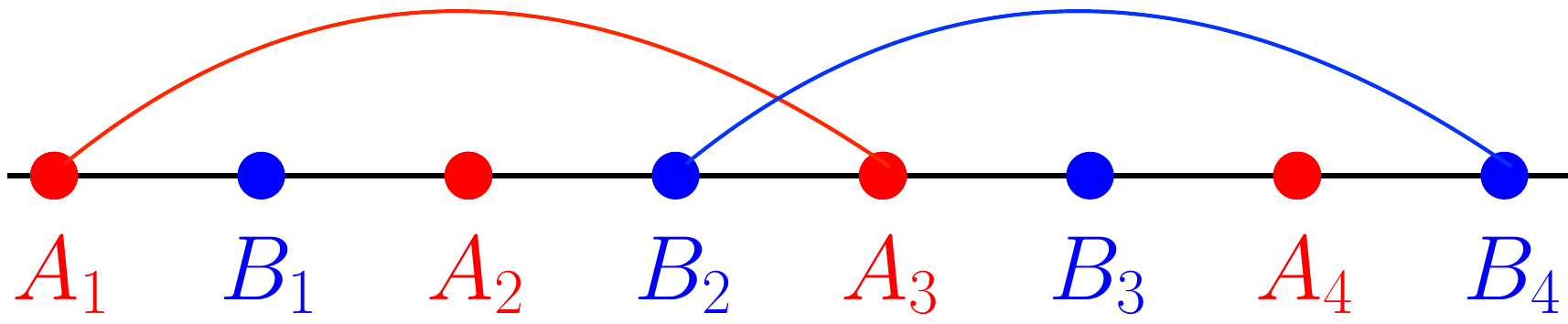}
\end{equation*} 
This pattern gives $\chi(A,B) =-1$, due to the odd number of intersections between two types of strings \footnote{This can be verified algebraically using $Z_{A,i}X_{B,i} Z_{A,i+1}  \ket{\psi}  = Z_{B,i}X_{A,i+1} Z_{B,i+1} \ket{\psi} = \ket{\psi}$.}.  On the other hand, $\chi(A,B)=1$  when there is no braiding between the two types of strings, e.g. the configuration with only $A_1, A_3,  B_3,  B_4$ equal $-1$ as shown below:

\begin{equation*}
\includegraphics[width=6cm]{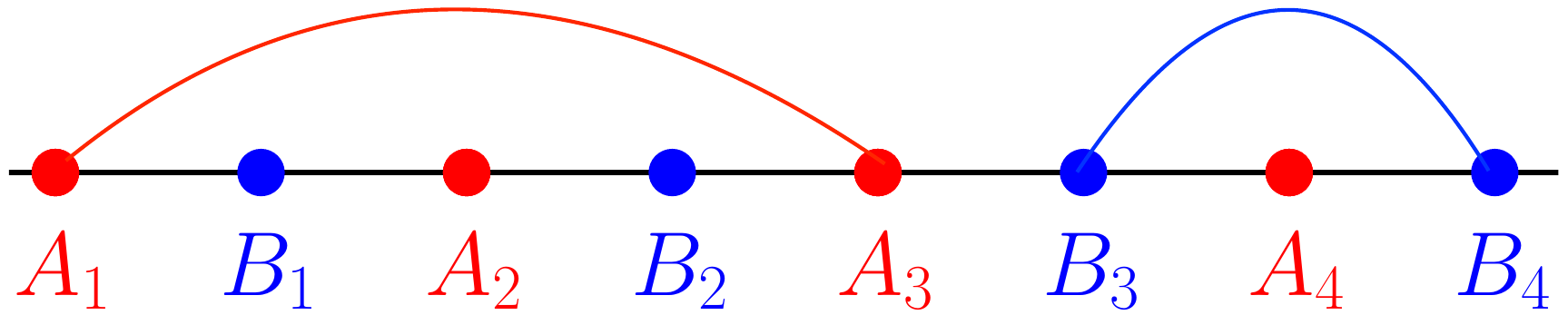}
\end{equation*}

Conceptually, $A =\{A_i\}$ and $B=\{B_i\}$ configurations represent the $e, m$ anyons (of the $\mathbb{Z}_2$ topological order) localized on the bipartition boundary, and the mutual braiding statistics is exactly captured by $\chi(A,B)$, i.e. the wave functions of the 1d cluster state with the $\mathbb{Z}_2 \times \mathbb{Z}_2$ SPT order.

By summing over the absolute values of the negativity spectrum (Eq.\ref{main:2dspectrum}), one can obtain the one-norm of the partially-transposed density matrix $\abs{\rho^{\Gamma}}_1  =  \frac{ \sum_{A,B} \abs{ \chi(A,B)e^{ \beta\sum_i A_i   }    }}{ \sum_{A,B} \chi(A,B)e^{ \beta\sum_i A_i   }  }
$, which turns out to be the ratio of two partition functions:

\begin{equation}
\abs{\rho^{\Gamma}}_1 = \frac{Z_{\text{1d,Ising}}}{\tilde{Z}_{\text{1d,Ising}}   }.
\end{equation}
$Z_{\text{1d,Ising}}= \sum_{\sigma} e^{\beta_z \sum_i \sigma_i \sigma_{i+1}}  $ is the partition function of the 1d Ising model at the inverse temperature $\beta_z  = - \frac{1}{2}\log (1-2p_z)$, and $\tilde{Z}_{\text{1d,Ising}}= \sum_{\sigma} e^{\beta_z \sum_i \sigma_i \sigma_{i+1}  } \prod_i \delta(\sigma_i \sigma_{i+1}  =1  ) = 2e^{\beta_z L}$ is the one without domain walls (i.e. excitations). Hence the entanglement negativity relates to the free energy difference associated with prohibiting domain walls: $E_N=  \log\abs{\rho^{\Gamma}}_1 = \log Z_{\text{1d,Ising}} - \log \tilde{Z}_{\text{1d,Ising}}$, which can be analytically computed:
\begin{equation}
E_N  =   \alpha L - E_{\text{topo}}  
\end{equation}
The area-law coefficient $\alpha =   \log(1+ e^{-2\beta_z}) $ captures the short-distance, non-universal part of the entanglement, and topological entanglement negativity $E_{\text{topo}} = \log 2 - \log \left(1+ (\tanh\beta_z )^L \right)$ captures the universal, long-range entanglement.

As a sanity check, at zero decoherence ($\beta_z=0$), one recovers the result that $\alpha=\log 2$ and $E_{\text{topo}} = \log 2$.  On the other hand, when increasing decoherence strength (i.e. increasing $\beta_z$), $\alpha$ decreases monotonically, consistent with the expectation that decoherence reduces entanglement. The subleading topological contribution $E_{\text{topo}}$ is more interesting; for any non-maximal decoherence ($p_z<\frac{1}{2}$, or equivalently, any non-infinite $\beta_z$), this quantity remains $\log 2$ in the thermodynamic limit $L\to \infty$, indicating that the long-range entanglement between the regions $\mA$ and $\mB$ is robust up to the maximal boundary decoherence.

As the negativity spectrum (Eq.\ref{eq:2d_spectrum}) is given by the wave function of the state $ \ket{\tilde{\psi}} \propto e^{\beta_z  \sum_i Z_{B,i}Z_{B,i+1} } \ket{\psi} $, i.e. the 1d cluster state perturbed by a weak projection $e^{ \beta_z \sum_i Z_{B,i}  Z_{ B,i+1}  }$, one might naturally wonder whether the persistence of the long-range entanglement relates to the robustness of SPT order in  $\ket{\tilde{\psi}}$. We find this is indeed the case. Specifically, the SPT order is characterized by two types of string order parameters: $S_A(i,j) = Z_{B,i}  X_{A,i+1} \cdots X_{A,j} Z_{B,j}$ and $S_B(i,j)= Z_{A,i}  X_{ B,i}\cdots X_{B,j-1} Z_{A,j}$ (see e.g. Ref.\cite{Pollmann_2012_string_order}). Since the perturbation commutes with $S_A$, one finds $ \ \bra{\tilde{\psi}} S_A(i,j) \ket{\tilde{\psi}} = 1 $, which is insensitive under the perturbation. On the other hand, $ \bra{\tilde{\psi}} S_B(i,j) \ket{\tilde{\psi}} =\frac{1}{ \left[\cosh2\beta_z \right]^2 \left[  1+ (\tanh2\beta_z)\right]^L  }$, which vanishes \textit{only} at the limit $\beta_z\to \infty$. Alternatively, one may consider the wavefunction overlap $ \bra{\tilde{\psi}}\ket{\tilde{\psi}} \propto \sum_\sigma e^{ 2\beta_z \sum_i \sigma_i \sigma_{i+1}}$ (see Appendix.\ref{append:SPT}), which is the partition function of the 1d Ising model, hence non-singular for any $\beta_z < \infty$. This indicates that the SPT order in $\ket{\tilde{\psi}}$ survives for any non-infinite $\beta_z$, consistent with the persistence of the long-range entanglement up to maximal decoherence. 

\textbf{\textit{3d toric code}} - Let's now discuss the entanglement structure of the 3d toric code subject to boundary decoherence. The Hamiltonian reads $H=-\sum_v  A_v - \sum_f B_f$ with $A_v =ZZZZZZ$ denoting the product of Pauli-Zs on 6 edges around a vertex $v$, and $B_f= XXXX$ denoting the product of Pauli-Xs on 4 edges on the boundary of a face $f$. By considering the Pauli-Z noise, which creates loop-like excitations, we will see that the entanglement negativity of the noisy mixed state relates to the 2d classical $\mathbb{Z}_2$ gauge theory, which does not exhibit a finite-temperature transition, indicating the persistence of long-range entanglement up to the maximal decoherence. On the other hand, considering the decoherence by Pauli-X noise, which creates point-like excitations, the entanglement negativity relates to the 2d classical Ising model, which exhibits a finite-temperature order-disorder transition. Correspondingly, there exists a disentangling transition in the decohered mixed state at a critical Pauli-X noise strength.

We will consider a 2d plane of size $L$ as the bipartition boundary that separates the two regions $\mA$, $\mB$ \footnote{Specifically, we consider a 3d lattice with the periodic boundary condition in $x$, $y$ directions and open boundary condition in $z$. The bipartition boundary is a non-contractible 2d plane spanned by $x$ and $y$ directions.}. There are $L^2$ boundary $A$ type operators on the vertices denoted by $A_v$ and $2L^2$ boundary $B$ type operators on edges denoted by $B_e$. 

Under the Pauli-Z decoherence with noise rate $p_z$ that only creates the excitations associated with the boundary $B_e$ operators, using a calculation similar to the 2d toric code, one can derive the negativity spectrum:

\begin{equation}\label{eq:3d_Z_spectrum}
\rho^{\Gamma} \propto  \bra{+} \prod_v Z_{A,v}^{\frac{1-A_v}{2}}  \prod_e Z_{B,e}^{\frac{1-B_e}{2}}   e^{\beta_z \sum_{v} \prod_{e \in \partial^* v}  Z_{B,e}   }\ket{\psi},
\end{equation}
where distinct eigenvalues are given by distinct configurations $A \equiv \{ A_v = \pm 1  \}$, $B \equiv \{ B_e = \pm 1  \}$, and $\beta_z  = - \frac{1}{2}\log (1-2p_z)$. We have introduced the fictitious Hilbert space of qubits living on vertices and edges.  $\ket{+}$ is the $+1$ Pauli-X product state, and $\ket{\psi}$ is a cluster state in 2d with the parent Hamiltonian $H= - \sum_{e}  X_{B,e} \prod_{v \in \partial e} Z_{A,v} - \sum_v X_{A,v} \prod_{e \in \partial^* v} Z_{B,e}$, where $\prod_{v \in \partial e}$ denotes the product over vertices on the boundary of $e$, and $\prod_{e \in \partial^* v}$ denotes the  product over edges on the coboundary of $v$ (i.e. the product of those edges whose boundaries contain $v$).  $\ket{\psi}$ exhibits an SPT order protected by a $\mathbb{Z}_2$ 0-form $\times$ $\mathbb{Z}_2$ 1-form symmetry generated by $\prod_v X_{A,v}$ and $\prod_{e\in\mathcal{C}} X_{B,e}$ with $\mathcal{C}$ being any closed loops.

With the condition  $X_{A,v}  \prod_{e  \in \partial^* v }Z_{B,e } \ket{\psi} = \ket{\psi} $, the negativity spectrum can be simplified as 

\begin{equation}
\rho^{\Gamma} = \frac{1}{Z} \chi(A,B)e^{\beta_z \sum_{v}A_v}
\end{equation}
with  
$\chi(A,B) = \frac{  \bra{+} \prod_v Z_{A,v}^{\frac{1-A_v}{2}}  \prod_e Z_{B,e}^{\frac{1-B_e}{2}}   \ket{\psi}}{\bra{+}   \ket{\psi}  }$, and $Z$ being a normalization constant. Using the negativity spectrum, we derive the entanglement negativity 
\begin{equation}
E_N  =  \log Z_{\text{2d,gauge}} - 
\log \tilde{  Z}_{\text{2d,gauge}} 
\end{equation}
where $ Z_{\text{2d,gauge}} = \sum_{\sigma} e^{\beta_z  \sum_v \prod_{e \in \partial^* v} \sigma_e}$ is the partition function of the classical 2d $\mathbb{Z}_2$ gauge theory, and $\tilde{Z}_{\text{2d,gauge}} = \sum_{\sigma} e^{\beta_z  \sum_v \prod_{ e\in \partial^* v} \sigma_e} \prod_v \delta(\prod_{e \in \partial^* v} \sigma_e=1)$ is the one prohibiting domain walls (excitations). 

By computing the partition function of the 2d gauge theory, we find $E_N= \alpha L^2 - E_{\text{topo}}$ with the area-law coefficient $\alpha = \log (1+e^{ -2\beta_z})$ and the topological entanglement negativity 

\begin{equation}
E_{\text{topo}} = \log 2 - \log \left(1+ (\tanh\beta_z )^{L^2} \right)
\end{equation}
This indicates that, in the thermodynamic limit $L\to \infty$, the long-range entanglement survives up to maximal decoherence ($\beta_z  \to \infty$, or equivalently, $p_z=\frac{1}{2}$). Consistent with the persistence of the long-range entanglement, the emergent boundary state $e^{\beta_z \sum_v  \prod_{e \in \partial^* v} Z_{B,e} } \ket{\psi}$ that describes the negativity spectrum (Eq.\ref{eq:3d_Z_spectrum}) exhibits an SPT order for any finite $\beta_z<\infty$, as diagnosed by the wave function overlap (Appendix.\ref{append:SPT}).

Now we consider the Pauli-X decoherence that creates the point-like excitations on vertices with noise strength $p_x$, and present the negativity spectrum 

\begin{equation}\label{eq:3d_spectrum_x}
\rho^{\Gamma} \propto  \bra{+} \prod_v Z_{A,v}^{\frac{1-A_v}{2}}  \prod_e Z_{B,e}^{\frac{1-B_e}{2}}   e^{\beta_x \sum_{e} \prod_{v\in \partial e}  Z_{A,v}   }\ket{\psi}
\end{equation}
which can further be simplified as  $\rho^{\Gamma} = \frac{1}{Z} \chi(A,B)e^{\beta_x \sum_{e}B_e}$ with $\beta_x = -\frac{1}{2} \log (1-2p_x)$, and $Z$ being a normalization constant. Using the spectrum, we derive the entanglement negativity 

\begin{equation}
E_N= \log Z_{\text{2d,Ising}} -  \log \tilde{Z}_{\text{2d,Ising}}.
\end{equation}
$ Z_{\text{2d,Ising}} = \sum_\sigma e^{\beta_x \sum_{\expval{ij} }  \sigma_i \sigma_j  }$ is the partition function of the classical 2d Ising model, and $\tilde{Z}_{\text{2d,Ising}}$ is the one when prohibiting excitations: $ \tilde{Z}_{\text{2d,Ising}}= \sum_\sigma e^{\beta_x \sum_{\expval{ij} 
 }  \sigma_i \sigma_j  } \prod_{\expval{ij}} \delta(\sigma_i \sigma_j =1) = 2 e^{2\beta_x L^2}$; the prefactor $2$ accounts for the two-fold ground-state degeneracy of the Ising model.

In the thermodynamic limit, $\log Z_{\text{2d,Ising}}$ takes the form $L^2 f_x + \log g_x$, where $f_x$ is the free-energy density of the 2d classical Ising model. Below the critical temperature $T_c= 1/\beta_c$, the Ising model spontaneously breaks the global $\mathbb{Z}_2$ symmetry, hence $g_x=2$. On the other hand, above the critical temperature, the global symmetry is not broken, giving $g_x=1$. Therefore, the entanglement negativity reads $E_N= \alpha L^2 - E_{\text{topo}  } $ with the area-law coefficient $\alpha  =  f_x -2\beta_x$, and topological entanglement negativity
\begin{equation}
E_{\text{topo}}=
\begin{cases}
&\log 2 ,  \text{ for } p_x<p_c \\
&0  ,  \text{ for }  p_x>p_c.
\end{cases}
\end{equation}
$p_c = \frac{1-e^{-2\beta_c}}{2}= 1-\frac{\sqrt{2}}{2} \approx 0.29$ with $\beta_c = \frac{1}{2} \ln (1+\sqrt{2})$ being the critical inverse temperature of the 2d Ising model \cite{onsager_1944}. As such, there exists a disentangling transition diagnosed by the discontinuity of topological entanglement negativity at the critical decoherence strength $p_c$. Relatedly,  $\ket{\tilde{\psi}} \propto e^{\beta_x \sum_{e} \prod_{v\in \partial e}  Z_{A,v}   }\ket{\psi}$ that describes the negativity spectrum (Eq.\ref{eq:3d_spectrum_x}) exhibit an SPT order up to a critical $\beta_x$ as diagnosed by the singularity of the wave-function overlap $ \bra{ \tilde{\psi}}  \ket{\tilde{\psi }}\propto \sum_{\sigma} e^{2\beta_x \sum_{\expval{ij}} \sigma_i \sigma_j}$, which is the partition function of the 2d Ising model at the inverse temperature $2\beta_x$. This indicates that the nature of the disentangling transition of the decohered mixed state and the SPT-order transition of the state $\ket{\tilde{\psi}}$ are both described by the 2d Ising universality, albeit at distinct critical temperatures.

\textbf{\textit{4d toric code}}  --- Finally, we explore the disentangling transition in the 4d toric code \cite{dennis2002},  where spins reside on each face (i.e. 2-cell) of a 4-dimensional hypercube. The Hamiltonian is $H= - \sum_e A_e - \sum_c B_c$, where $A_e$ is the product of 6 Pauli-Z operators on the faces adjacent to the edge $e$, and $B_c$ is the product of 6 Pauli-X operators on the faces around the boundary of the cube $c$. The bipartition boundary of a 4d hypercube is a 3d lattice, where the boundary stabilizers are $A_e$ living on edges and $B_f$ living on faces \footnote{Specifically, we consider a 4d lattice with the periodic boundary condition in $x$, $y$, $z$ directions and open boundary condition in $w$ direction. The bipartition boundary is a non-contractible 3d cube spanned by $x,y,z$ directions.}. 

Under the Pauli-Z noise that creates the excitations associated with the boundary $B_f$ stabilizers with noise rate $p_z$ \footnote{The Pauli-X noise will lead to the same mixed-state entanglement structure due to the duality between $A_e, B_c$ stabilizers.}, the  negativity spectrum of the decohered mixed state reads 
\begin{equation}
\rho^\Gamma  \propto \bra{+} \prod_e Z_{A,e}^{\frac{1-A_e}{2}} \prod_fZ_{B,f}^{\frac{1-B_f}{2}} e^{\beta_z\sum_e \prod_{f  \in  \partial^* e  }   Z_{B,f}}  \ket{\psi},
\end{equation}
where $\beta_z  \equiv  - \frac{1}{2}\log (1-2p_z)$, and we introduce the fictitious Hilbert space of qubits living on edges and faces; $\ket{+}$ is the $+1$ X-basis product state, and $\ket{\psi}$ is the ground state of the 3d cluster-state Hamiltonian $H=-  \sum_{f } X_{B,f} \prod_{l\in \partial f  }Z_{A,l}  -  \sum_{ l } X_{A,l} \prod_{ f  \in \partial^* l   }Z_{B,f}$\footnote{Here we again use $\partial$ and $\partial^*$ to denote the boundary map and the coboundary map, respectively.} \cite{3d_cluster_state_2005}. $\ket{\psi}$ exhibits an SPT order protected by the $\mathbb{Z}_{2}$ 1-form $\cross \mathbb{Z}_{2}$ 1-form symmetry, generated by $\prod_{f \in\partial c  }X_f$ and $\prod_{l  \in \partial^* v   }X_l$.

The negativity spectrum can be further simplified as 
\begin{equation}
\rho^{\Gamma} = \frac{1}{Z} \chi(A,B) e^{\beta_z \sum_e  A_e}
\end{equation}
with  $\chi(A,B) =  \frac{  \bra{+} \prod_e Z_{A,e}^{\frac{1-A_e}{2}} \prod_fZ_{B,f}^{\frac{1-B_f}{2}}  \ket{\psi},  }{\bra{+}\ket{\psi}}$. Due to the two $\mathbb{Z}_2$ 1-form symmetries, to have non-vanishing $\chi(A,B)$, $A$ and $B$ must form closed loops in the primary lattice and the dual lattice respectively, and $\chi(A,B)= 1, -1$ for even, odd number of times of braiding (linking) between the $A$ loops and $B$ loops.  Conceptually, $A$ and $B$ can be thought of as two species of loop excitations of the 4d toric code living on the 3d boundary, and the mutual braiding statistics is encoded in $\chi(A,B)$, i.e. the wave functions of 3d cluster-state SPT in the X basis.

With the negativity spectrum, we derive the entanglement negativity 
\begin{equation}
E_N =  \log Z_{\text{3d,gauge}} - \log\tilde{Z}_{\text{3d,gauge}},   
\end{equation}
where $Z_{\text{3d,gauge}}= \sum_\sigma e^{\beta_z \sum_f \prod_{ e \in \partial f} \sigma_e }$ is the partition function of the 3d classical $\mathbb{Z}_2$ gauge theory, and  $\tilde{Z}_{\text{3d,gauge}}$ is the same theory but prohibiting excitations: $\tilde{Z}_{\text{3d,gauge}}= \sum_\sigma e^{\beta_z \sum_f \prod_{e \in \partial f} \sigma_e }\prod_f \delta(\prod_{e \in \partial f} \sigma_e =1)=ne^{3\beta_z L^3 }$, with $n=2^{L^3+2}$ being the number of ground-state configurations of the 3d gauge theory. On the other hand, $\log Z_{\text{3d,gauge}}$ exhibits a confinement-deconfinement transition at a critical temperature $T_c = 1/\beta_c$ \cite{wegner_1971,1979_kogut}, and is expected to take the following form in the thermodynamic limit: $L^3 f_z + \log g $. $f_z$ is the free-energy density, $g=1$ in the confined phase ($\beta_z< \beta_c$), and $g=4$ in the deconfined phase  ($\beta_z> \beta_c$). One way to understand the discontinuity in $g$ is by mapping the 3d classical $\mathbb{Z}_2$ gauge theory at finite temperature to the 2d quantum $\mathbb{Z}_2$ gauge theory at zero temperature. The quantity $g$ can then be understood as the ground-state degeneracy in the 2d quantum gauge theory; $g =4 $ in the deconfined phase due to the spontaneous 1-form symmetry breaking, and $g=1$ in the confined phase due to the restoration of the 1-form symmetry (see Appendix.\ref{append:qc_mapping} for detailed discussion). Therefore, there exists a disentangling transition diagnosed by the discontinuity of topological entanglement negativity: $E_{\text{topo}}= 2\log 2, 0$ for $p_z< p_c$ and $p_z >p_c$ respectively, where  $p_c=\frac{1-e^{-2 \beta_c  }}{2}$ is critical noise rate, with $\beta_c$ being the critical inverse temperature of the 3d classical $\mathbb{Z}_2$ gauge theory.

\textbf{\textit{Discussion}} - In this work, we present the exact results for the negativity spectrum and the entanglement structure of topological orders subject to decoherence along the bipartition boundary. The negativity spectrum of the decohered mixed state in $d$ space dimensions corresponds to the wave functions of an SPT order in $d-1$ space dimensions under certain symmetry-preserving perturbations. This understanding allows us to compute entanglement negativity (without resorting to a replica trick) and map it to the free energy of a \textit{translationally-invariant} classical statistical mechanics model in $d-1$ space dimensions, whose phase transition corresponds to the disentangling transition in the physical systems. While we only discuss the entanglement structure of the toric code, our approach can be straightforwardly applied to any topological qubit-stabilizer models, e.g. X-cube fracton code \cite{xcube} or Haah's code \cite{Haah}.

As a future question, it would be interesting to explore boundary decoherence from the perspective of separability transitions \cite{grover_2023_separability_toric}. Specifically, one can ask: whether a topological order subject to boundary decoherence may exhibit a boundary separability transition at a critical decoherence strength $p_c$, above (below) which the mixed state $\rho$ can (cannot) be written as $\rho= \sum_i p_i \ket{\psi_i} \bra{\psi_i}$, where each $\ket{\psi_i}$ can be made unentangled w.r.t a fixed $\mA, \mB$ bipartition via a finite-depth local unitary circuit $U_{\text{fd},i}$. Namely, $\ket{\psi_i}= U_{\text{fd},i} \ket{\psi_{\mA,i}}_{\mA}\otimes \ket{\psi_{\mB,i}}_{\mB}$, with $\ket{\psi_{\mA,i}}_{\mA}, \ket{\psi_{\mB,i}}_{\mB}$ being arbitrary states in $\mA, \mB$ that may exhibit a bulk topological order. If such a boundary separability transition exists, does it coincide with the disentangling transition witnessed via topological entanglement negativity?

It would also be interesting to explore the connection between the disentangling transition and the channel definition of mixed-state phases of matter \cite{sang2023mixed}. For instance, in the 2d toric code subject to the boundary decoherence, we found that topological entanglement negativity survives up to maximal decoherence. Does it suggest the un-decohered toric code can be reconstructed from the boundary-decohered toric code via a finite-depth quantum channel acting on the bipartition boundary?

More technically, there are two cases where the mixed-state entanglement structure of the decohered toric code remains elusive: (i) the decoherence given by simultaneous Pauli-X and Pauli-Z noise on the bipartition boundary, and (ii) the decoherence given by Pauli-X or Pauli-Z noise channels uniformly throughout the entire system. For the first case, while we present partial progress by deriving the negativity spectrum in Appendix.\ref{append:xz_both}, we are unable to obtain exact results of entanglement negativity. For the second case, while Ref.\cite{fan_2023_toric} derived the moment of the partially-transposed matrix $\tr (\rho^{\Gamma})^{2n}$ for integer $n$, it is unclear how to perform the analytic continuation by taking $n\to \frac{1}{2}$ to obtain the exact results of entanglement negativity.

Finally, our approach based on the negativity spectrum and an emergent SPT description may be useful to explore the entanglement structure of intrinsically mixed-state topological orders \cite{wang2023intrinsic,sohal2024noisy}, an intriguing type of topological order that may not be realized in gapped ground states of local Hamiltonians. More generally, it might be fruitful to consider various non-trivial states of matter, e.g. non-abelian topological order, quantum critical states, and chiral topological phases, subject to the decoherence on the bipartition boundary, and explore the entanglement structure of mixed states and their disentangling transitions.

\acknowledgements{T.-C.L. thanks Zhehao Zhang and Ruihua Fan for helpful comments on the manuscript. T.-C.L. also thanks Claudio Chamon for hosting his visit at Boston University, during which this work is completed. The research at Perimeter Institute is supported in part by the Government of Canada through the Department of Innovation, Science and Economic Development Canada and by the Province of Ontario through the Ministry of Colleges and Universities. 
	}
	
	\bibliography{v1bib}

\begin{thebibliography}{50}%
\makeatletter
\providecommand \@ifxundefined [1]{%
 \@ifx{#1\undefined}
}%
\providecommand \@ifnum [1]{%
 \ifnum #1\expandafter \@firstoftwo
 \else \expandafter \@secondoftwo
 \fi
}%
\providecommand \@ifx [1]{%
 \ifx #1\expandafter \@firstoftwo
 \else \expandafter \@secondoftwo
 \fi
}%
\providecommand \natexlab [1]{#1}%
\providecommand \enquote  [1]{``#1''}%
\providecommand \bibnamefont  [1]{#1}%
\providecommand \bibfnamefont [1]{#1}%
\providecommand \citenamefont [1]{#1}%
\providecommand \href@noop [0]{\@secondoftwo}%
\providecommand \href [0]{\begingroup \@sanitize@url \@href}%
\providecommand \@href[1]{\@@startlink{#1}\@@href}%
\providecommand \@@href[1]{\endgroup#1\@@endlink}%
\providecommand \@sanitize@url [0]{\catcode `\\12\catcode `\$12\catcode `\&12\catcode `\#12\catcode `\^12\catcode `\_12\catcode `\%12\relax}%
\providecommand \@@startlink[1]{}%
\providecommand \@@endlink[0]{}%
\providecommand \url  [0]{\begingroup\@sanitize@url \@url }%
\providecommand \@url [1]{\endgroup\@href {#1}{\urlprefix }}%
\providecommand \urlprefix  [0]{URL }%
\providecommand \Eprint [0]{\href }%
\providecommand \doibase [0]{https://doi.org/}%
\providecommand \selectlanguage [0]{\@gobble}%
\providecommand \bibinfo  [0]{\@secondoftwo}%
\providecommand \bibfield  [0]{\@secondoftwo}%
\providecommand \translation [1]{[#1]}%
\providecommand \BibitemOpen [0]{}%
\providecommand \bibitemStop [0]{}%
\providecommand \bibitemNoStop [0]{.\EOS\space}%
\providecommand \EOS [0]{\spacefactor3000\relax}%
\providecommand \BibitemShut  [1]{\csname bibitem#1\endcsname}%
\let\auto@bib@innerbib\@empty
\bibitem [{\citenamefont {Wen}(2004)}]{wen2004quantum}%
  \BibitemOpen
  \bibfield  {author} {\bibinfo {author} {\bibfnamefont {X.-G.}\ \bibnamefont {Wen}},\ }\bibfield  {title} {\bibinfo {title} {Quantum field theory of many-body systems: from the origin of sound to an origin of light and electrons},\ }\href {https://academic.oup.com/book/25836?searchresult=1} {\bibfield  {journal} {\bibinfo  {journal} {OUP Oxford}\ } (\bibinfo {year} {2004})}\BibitemShut {NoStop}%
\bibitem [{\citenamefont {Kitaev}(2003)}]{kitaev2003fault}%
  \BibitemOpen
  \bibfield  {author} {\bibinfo {author} {\bibfnamefont {A.}~\bibnamefont {Kitaev}},\ }\bibfield  {title} {\bibinfo {title} {Fault-tolerant quantum computation by anyons},\ }\href {https://doi.org/https://doi.org/10.1016/S0003-4916(02)00018-0} {\bibfield  {journal} {\bibinfo  {journal} {Annals of Physics}\ }\textbf {\bibinfo {volume} {303}},\ \bibinfo {pages} {2} (\bibinfo {year} {2003})}\BibitemShut {NoStop}%
\bibitem [{\citenamefont {Dennis}\ \emph {et~al.}(2002)\citenamefont {Dennis}, \citenamefont {Kitaev}, \citenamefont {Landahl},\ and\ \citenamefont {Preskill}}]{dennis2002}%
  \BibitemOpen
  \bibfield  {author} {\bibinfo {author} {\bibfnamefont {E.}~\bibnamefont {Dennis}}, \bibinfo {author} {\bibfnamefont {A.}~\bibnamefont {Kitaev}}, \bibinfo {author} {\bibfnamefont {A.}~\bibnamefont {Landahl}},\ and\ \bibinfo {author} {\bibfnamefont {J.}~\bibnamefont {Preskill}},\ }\bibfield  {title} {\bibinfo {title} {Topological quantum memory},\ }\href {https://doi.org/10.1063/1.1499754} {\bibfield  {journal} {\bibinfo  {journal} {Journal of Mathematical Physics}\ }\textbf {\bibinfo {volume} {43}},\ \bibinfo {pages} {4452} (\bibinfo {year} {2002})}\BibitemShut {NoStop}%
\bibitem [{\citenamefont {Fan}\ \emph {et~al.}(2023)\citenamefont {Fan}, \citenamefont {Bao}, \citenamefont {Altman},\ and\ \citenamefont {Vishwanath}}]{fan_2023_toric}%
  \BibitemOpen
  \bibfield  {author} {\bibinfo {author} {\bibfnamefont {R.}~\bibnamefont {Fan}}, \bibinfo {author} {\bibfnamefont {Y.}~\bibnamefont {Bao}}, \bibinfo {author} {\bibfnamefont {E.}~\bibnamefont {Altman}},\ and\ \bibinfo {author} {\bibfnamefont {A.}~\bibnamefont {Vishwanath}},\ }\bibfield  {title} {\bibinfo {title} {Diagnostics of mixed-state topological order and breakdown of quantum memory},\ }\href@noop {} {\bibfield  {journal} {\bibinfo  {journal} {arXiv preprint arXiv:2301.05689}\ } (\bibinfo {year} {2023})}\BibitemShut {NoStop}%
\bibitem [{\citenamefont {de~Groot}\ \emph {et~al.}(2022)\citenamefont {de~Groot}, \citenamefont {Turzillo},\ and\ \citenamefont {Schuch}}]{spt_Schuch_2022}%
  \BibitemOpen
  \bibfield  {author} {\bibinfo {author} {\bibfnamefont {C.}~\bibnamefont {de~Groot}}, \bibinfo {author} {\bibfnamefont {A.}~\bibnamefont {Turzillo}},\ and\ \bibinfo {author} {\bibfnamefont {N.}~\bibnamefont {Schuch}},\ }\bibfield  {title} {\bibinfo {title} {Symmetry {P}rotected {T}opological {O}rder in {O}pen {Q}uantum {S}ystems},\ }\href {https://doi.org/10.22331/q-2022-11-10-856} {\bibfield  {journal} {\bibinfo  {journal} {{Quantum}}\ }\textbf {\bibinfo {volume} {6}},\ \bibinfo {pages} {856} (\bibinfo {year} {2022})}\BibitemShut {NoStop}%
\bibitem [{\citenamefont {Ma}\ and\ \citenamefont {Wang}(2023)}]{wang_spt_2023}%
  \BibitemOpen
  \bibfield  {author} {\bibinfo {author} {\bibfnamefont {R.}~\bibnamefont {Ma}}\ and\ \bibinfo {author} {\bibfnamefont {C.}~\bibnamefont {Wang}},\ }\bibfield  {title} {\bibinfo {title} {Average symmetry-protected topological phases},\ }\href {https://doi.org/10.1103/PhysRevX.13.031016} {\bibfield  {journal} {\bibinfo  {journal} {Phys. Rev. X}\ }\textbf {\bibinfo {volume} {13}},\ \bibinfo {pages} {031016} (\bibinfo {year} {2023})}\BibitemShut {NoStop}%
\bibitem [{\citenamefont {Lee}\ \emph {et~al.}(2022)\citenamefont {Lee}, \citenamefont {You},\ and\ \citenamefont {Xu}}]{lee2022_spt}%
  \BibitemOpen
  \bibfield  {author} {\bibinfo {author} {\bibfnamefont {J.~Y.}\ \bibnamefont {Lee}}, \bibinfo {author} {\bibfnamefont {Y.-Z.}\ \bibnamefont {You}},\ and\ \bibinfo {author} {\bibfnamefont {C.}~\bibnamefont {Xu}},\ }\bibfield  {title} {\bibinfo {title} {Symmetry protected topological phases under decoherence},\ }\href@noop {} {\bibfield  {journal} {\bibinfo  {journal} {arXiv preprint arXiv:2210.16323}\ } (\bibinfo {year} {2022})}\BibitemShut {NoStop}%
\bibitem [{\citenamefont {Lee}\ \emph {et~al.}(2023)\citenamefont {Lee}, \citenamefont {Jian},\ and\ \citenamefont {Xu}}]{criticality_2023_xu}%
  \BibitemOpen
  \bibfield  {author} {\bibinfo {author} {\bibfnamefont {J.~Y.}\ \bibnamefont {Lee}}, \bibinfo {author} {\bibfnamefont {C.-M.}\ \bibnamefont {Jian}},\ and\ \bibinfo {author} {\bibfnamefont {C.}~\bibnamefont {Xu}},\ }\bibfield  {title} {\bibinfo {title} {Quantum criticality under decoherence or weak measurement},\ }\href {https://doi.org/10.1103/PRXQuantum.4.030317} {\bibfield  {journal} {\bibinfo  {journal} {PRX Quantum}\ }\textbf {\bibinfo {volume} {4}},\ \bibinfo {pages} {030317} (\bibinfo {year} {2023})}\BibitemShut {NoStop}%
\bibitem [{\citenamefont {Bao}\ \emph {et~al.}(2023)\citenamefont {Bao}, \citenamefont {Fan}, \citenamefont {Vishwanath},\ and\ \citenamefont {Altman}}]{bao2023mixed}%
  \BibitemOpen
  \bibfield  {author} {\bibinfo {author} {\bibfnamefont {Y.}~\bibnamefont {Bao}}, \bibinfo {author} {\bibfnamefont {R.}~\bibnamefont {Fan}}, \bibinfo {author} {\bibfnamefont {A.}~\bibnamefont {Vishwanath}},\ and\ \bibinfo {author} {\bibfnamefont {E.}~\bibnamefont {Altman}},\ }\bibfield  {title} {\bibinfo {title} {Mixed-state topological order and the errorfield double formulation of decoherence-induced transitions},\ }\href@noop {} {\bibfield  {journal} {\bibinfo  {journal} {arXiv preprint arXiv:2301.05687}\ } (\bibinfo {year} {2023})}\BibitemShut {NoStop}%
\bibitem [{\citenamefont {Lu}\ \emph {et~al.}(2023)\citenamefont {Lu}, \citenamefont {Zhang}, \citenamefont {Vijay},\ and\ \citenamefont {Hsieh}}]{Lu_mixed_feedback_2023}%
  \BibitemOpen
  \bibfield  {author} {\bibinfo {author} {\bibfnamefont {T.-C.}\ \bibnamefont {Lu}}, \bibinfo {author} {\bibfnamefont {Z.}~\bibnamefont {Zhang}}, \bibinfo {author} {\bibfnamefont {S.}~\bibnamefont {Vijay}},\ and\ \bibinfo {author} {\bibfnamefont {T.~H.}\ \bibnamefont {Hsieh}},\ }\bibfield  {title} {\bibinfo {title} {Mixed-state long-range order and criticality from measurement and feedback},\ }\href {https://doi.org/10.1103/PRXQuantum.4.030318} {\bibfield  {journal} {\bibinfo  {journal} {PRX Quantum}\ }\textbf {\bibinfo {volume} {4}},\ \bibinfo {pages} {030318} (\bibinfo {year} {2023})}\BibitemShut {NoStop}%
\bibitem [{\citenamefont {Zou}\ \emph {et~al.}(2023)\citenamefont {Zou}, \citenamefont {Sang},\ and\ \citenamefont {Hsieh}}]{hsieh_2023_criticality}%
  \BibitemOpen
  \bibfield  {author} {\bibinfo {author} {\bibfnamefont {Y.}~\bibnamefont {Zou}}, \bibinfo {author} {\bibfnamefont {S.}~\bibnamefont {Sang}},\ and\ \bibinfo {author} {\bibfnamefont {T.~H.}\ \bibnamefont {Hsieh}},\ }\bibfield  {title} {\bibinfo {title} {Channeling quantum criticality},\ }\href {https://doi.org/10.1103/PhysRevLett.130.250403} {\bibfield  {journal} {\bibinfo  {journal} {Phys. Rev. Lett.}\ }\textbf {\bibinfo {volume} {130}},\ \bibinfo {pages} {250403} (\bibinfo {year} {2023})}\BibitemShut {NoStop}%
\bibitem [{\citenamefont {Chen}\ and\ \citenamefont {Grover}(2023{\natexlab{a}})}]{grover_2023_separability_toric}%
  \BibitemOpen
  \bibfield  {author} {\bibinfo {author} {\bibfnamefont {Y.-H.}\ \bibnamefont {Chen}}\ and\ \bibinfo {author} {\bibfnamefont {T.}~\bibnamefont {Grover}},\ }\bibfield  {title} {\bibinfo {title} {Separability transitions in topological states induced by local decoherence},\ }\href@noop {} {\bibfield  {journal} {\bibinfo  {journal} {arXiv preprint arXiv:2309.11879}\ } (\bibinfo {year} {2023}{\natexlab{a}})}\BibitemShut {NoStop}%
\bibitem [{\citenamefont {Chen}\ and\ \citenamefont {Grover}(2023{\natexlab{b}})}]{chen2023symmetry}%
  \BibitemOpen
  \bibfield  {author} {\bibinfo {author} {\bibfnamefont {Y.-H.}\ \bibnamefont {Chen}}\ and\ \bibinfo {author} {\bibfnamefont {T.}~\bibnamefont {Grover}},\ }\bibfield  {title} {\bibinfo {title} {Symmetry-enforced many-body separability transitions},\ }\href@noop {} {\bibfield  {journal} {\bibinfo  {journal} {arXiv preprint arXiv:2310.07286}\ } (\bibinfo {year} {2023}{\natexlab{b}})}\BibitemShut {NoStop}%
\bibitem [{\citenamefont {Su}\ \emph {et~al.}(2023)\citenamefont {Su}, \citenamefont {Myerson-Jain}, \citenamefont {Wang}, \citenamefont {Jian},\ and\ \citenamefont {Xu}}]{su2023higher}%
  \BibitemOpen
  \bibfield  {author} {\bibinfo {author} {\bibfnamefont {K.}~\bibnamefont {Su}}, \bibinfo {author} {\bibfnamefont {N.}~\bibnamefont {Myerson-Jain}}, \bibinfo {author} {\bibfnamefont {C.}~\bibnamefont {Wang}}, \bibinfo {author} {\bibfnamefont {C.-M.}\ \bibnamefont {Jian}},\ and\ \bibinfo {author} {\bibfnamefont {C.}~\bibnamefont {Xu}},\ }\bibfield  {title} {\bibinfo {title} {Higher-form symmetries under weak measurement},\ }\href@noop {} {\bibfield  {journal} {\bibinfo  {journal} {arXiv preprint arXiv:2304.14433}\ } (\bibinfo {year} {2023})}\BibitemShut {NoStop}%
\bibitem [{\citenamefont {Sang}\ \emph {et~al.}(2023)\citenamefont {Sang}, \citenamefont {Zou},\ and\ \citenamefont {Hsieh}}]{sang2023mixed}%
  \BibitemOpen
  \bibfield  {author} {\bibinfo {author} {\bibfnamefont {S.}~\bibnamefont {Sang}}, \bibinfo {author} {\bibfnamefont {Y.}~\bibnamefont {Zou}},\ and\ \bibinfo {author} {\bibfnamefont {T.~H.}\ \bibnamefont {Hsieh}},\ }\bibfield  {title} {\bibinfo {title} {Mixed-state quantum phases: Renormalization and quantum error correction},\ }\href@noop {} {\bibfield  {journal} {\bibinfo  {journal} {arXiv preprint arXiv:2310.08639}\ } (\bibinfo {year} {2023})}\BibitemShut {NoStop}%
\bibitem [{\citenamefont {Wang}\ \emph {et~al.}(2023)\citenamefont {Wang}, \citenamefont {Wu},\ and\ \citenamefont {Wang}}]{wang2023intrinsic}%
  \BibitemOpen
  \bibfield  {author} {\bibinfo {author} {\bibfnamefont {Z.}~\bibnamefont {Wang}}, \bibinfo {author} {\bibfnamefont {Z.}~\bibnamefont {Wu}},\ and\ \bibinfo {author} {\bibfnamefont {Z.}~\bibnamefont {Wang}},\ }\bibfield  {title} {\bibinfo {title} {Intrinsic mixed-state topological order without quantum memory},\ }\href@noop {} {\bibfield  {journal} {\bibinfo  {journal} {arXiv preprint arXiv:2307.13758}\ } (\bibinfo {year} {2023})}\BibitemShut {NoStop}%
\bibitem [{\citenamefont {Ma}\ \emph {et~al.}(2023)\citenamefont {Ma}, \citenamefont {Zhang}, \citenamefont {Bi}, \citenamefont {Cheng},\ and\ \citenamefont {Wang}}]{ma_2023topological_mixed}%
  \BibitemOpen
  \bibfield  {author} {\bibinfo {author} {\bibfnamefont {R.}~\bibnamefont {Ma}}, \bibinfo {author} {\bibfnamefont {J.-H.}\ \bibnamefont {Zhang}}, \bibinfo {author} {\bibfnamefont {Z.}~\bibnamefont {Bi}}, \bibinfo {author} {\bibfnamefont {M.}~\bibnamefont {Cheng}},\ and\ \bibinfo {author} {\bibfnamefont {C.}~\bibnamefont {Wang}},\ }\bibfield  {title} {\bibinfo {title} {Topological phases with average symmetries: the decohered, the disordered, and the intrinsic},\ }\href@noop {} {\bibfield  {journal} {\bibinfo  {journal} {arXiv preprint arXiv:2305.16399}\ } (\bibinfo {year} {2023})}\BibitemShut {NoStop}%
\bibitem [{\citenamefont {Hsin}\ \emph {et~al.}(2023)\citenamefont {Hsin}, \citenamefont {Luo},\ and\ \citenamefont {Sun}}]{hsin2023anomalies}%
  \BibitemOpen
  \bibfield  {author} {\bibinfo {author} {\bibfnamefont {P.-S.}\ \bibnamefont {Hsin}}, \bibinfo {author} {\bibfnamefont {Z.-X.}\ \bibnamefont {Luo}},\ and\ \bibinfo {author} {\bibfnamefont {H.-Y.}\ \bibnamefont {Sun}},\ }\bibfield  {title} {\bibinfo {title} {Anomalies of average symmetries: Entanglement and open quantum systems},\ }\href@noop {} {\bibfield  {journal} {\bibinfo  {journal} {arXiv preprint arXiv:2312.09074}\ } (\bibinfo {year} {2023})}\BibitemShut {NoStop}%
\bibitem [{\citenamefont {Su}\ \emph {et~al.}(2024{\natexlab{a}})\citenamefont {Su}, \citenamefont {Myerson-Jain},\ and\ \citenamefont {Xu}}]{xu_chern_2024}%
  \BibitemOpen
  \bibfield  {author} {\bibinfo {author} {\bibfnamefont {K.}~\bibnamefont {Su}}, \bibinfo {author} {\bibfnamefont {N.}~\bibnamefont {Myerson-Jain}},\ and\ \bibinfo {author} {\bibfnamefont {C.}~\bibnamefont {Xu}},\ }\bibfield  {title} {\bibinfo {title} {Conformal field theories generated by chern insulators under decoherence or measurement},\ }\href {https://doi.org/10.1103/PhysRevB.109.035146} {\bibfield  {journal} {\bibinfo  {journal} {Phys. Rev. B}\ }\textbf {\bibinfo {volume} {109}},\ \bibinfo {pages} {035146} (\bibinfo {year} {2024}{\natexlab{a}})}\BibitemShut {NoStop}%
\bibitem [{\citenamefont {Myerson-Jain}\ \emph {et~al.}(2023)\citenamefont {Myerson-Jain}, \citenamefont {Hughes},\ and\ \citenamefont {Xu}}]{xu_2024_anyon_decoherence}%
  \BibitemOpen
  \bibfield  {author} {\bibinfo {author} {\bibfnamefont {N.}~\bibnamefont {Myerson-Jain}}, \bibinfo {author} {\bibfnamefont {T.~L.}\ \bibnamefont {Hughes}},\ and\ \bibinfo {author} {\bibfnamefont {C.}~\bibnamefont {Xu}},\ }\bibfield  {title} {\bibinfo {title} {Decoherence through ancilla anyon reservoirs},\ }\href@noop {} {\bibfield  {journal} {\bibinfo  {journal} {arXiv preprint arXiv:2312.04638}\ } (\bibinfo {year} {2023})}\BibitemShut {NoStop}%
\bibitem [{\citenamefont {Li}\ and\ \citenamefont {Mong}(2024)}]{mong_2024_replica}%
  \BibitemOpen
  \bibfield  {author} {\bibinfo {author} {\bibfnamefont {Z.}~\bibnamefont {Li}}\ and\ \bibinfo {author} {\bibfnamefont {R.~S.}\ \bibnamefont {Mong}},\ }\bibfield  {title} {\bibinfo {title} {Replica topological order in quantum mixed states and quantum error correction},\ }\href@noop {} {\bibfield  {journal} {\bibinfo  {journal} {arXiv preprint arXiv:2402.09516}\ } (\bibinfo {year} {2024})}\BibitemShut {NoStop}%
\bibitem [{\citenamefont {Su}\ \emph {et~al.}(2024{\natexlab{b}})\citenamefont {Su}, \citenamefont {Yang},\ and\ \citenamefont {Jian}}]{jian_2024_duality}%
  \BibitemOpen
  \bibfield  {author} {\bibinfo {author} {\bibfnamefont {K.}~\bibnamefont {Su}}, \bibinfo {author} {\bibfnamefont {Z.}~\bibnamefont {Yang}},\ and\ \bibinfo {author} {\bibfnamefont {C.-M.}\ \bibnamefont {Jian}},\ }\bibfield  {title} {\bibinfo {title} {Tapestry of dualities in decohered quantum error correction codes},\ }\href@noop {} {\bibfield  {journal} {\bibinfo  {journal} {arXiv preprint arXiv:2401.17359}\ } (\bibinfo {year} {2024}{\natexlab{b}})}\BibitemShut {NoStop}%
\bibitem [{\citenamefont {Lyons}(2024)}]{lyons2024understanding}%
  \BibitemOpen
  \bibfield  {author} {\bibinfo {author} {\bibfnamefont {A.}~\bibnamefont {Lyons}},\ }\bibfield  {title} {\bibinfo {title} {Understanding stabilizer codes under local decoherence through a general statistical mechanics mapping},\ }\href@noop {} {\bibfield  {journal} {\bibinfo  {journal} {arXiv preprint arXiv:2403.03955}\ } (\bibinfo {year} {2024})}\BibitemShut {NoStop}%
\bibitem [{\citenamefont {Lee}(2024)}]{lee2024exact}%
  \BibitemOpen
  \bibfield  {author} {\bibinfo {author} {\bibfnamefont {J.~Y.}\ \bibnamefont {Lee}},\ }\bibfield  {title} {\bibinfo {title} {Exact calculations of coherent information for toric codes under decoherence: Identifying the fundamental error threshold},\ }\href@noop {} {\bibfield  {journal} {\bibinfo  {journal} {arXiv preprint arXiv:2402.16937}\ } (\bibinfo {year} {2024})}\BibitemShut {NoStop}%
\bibitem [{\citenamefont {Chen}\ and\ \citenamefont {Grover}(2024)}]{chen2024unconventional}%
  \BibitemOpen
  \bibfield  {author} {\bibinfo {author} {\bibfnamefont {Y.-H.}\ \bibnamefont {Chen}}\ and\ \bibinfo {author} {\bibfnamefont {T.}~\bibnamefont {Grover}},\ }\bibfield  {title} {\bibinfo {title} {Unconventional topological mixed-state transition and critical phase induced by self-dual coherent errors},\ }\href@noop {} {\bibfield  {journal} {\bibinfo  {journal} {arXiv preprint arXiv:2403.06553}\ } (\bibinfo {year} {2024})}\BibitemShut {NoStop}%
\bibitem [{\citenamefont {Guo}\ \emph {et~al.}(2024)\citenamefont {Guo}, \citenamefont {Zhang}, \citenamefont {Yang},\ and\ \citenamefont {Bi}}]{Zhen_2024_spt}%
  \BibitemOpen
  \bibfield  {author} {\bibinfo {author} {\bibfnamefont {Y.}~\bibnamefont {Guo}}, \bibinfo {author} {\bibfnamefont {J.-H.}\ \bibnamefont {Zhang}}, \bibinfo {author} {\bibfnamefont {S.}~\bibnamefont {Yang}},\ and\ \bibinfo {author} {\bibfnamefont {Z.}~\bibnamefont {Bi}},\ }\bibfield  {title} {\bibinfo {title} {Locally purified density operators for symmetry-protected topological phases in mixed states},\ }\href@noop {} {\bibfield  {journal} {\bibinfo  {journal} {arXiv preprint arXiv:2403.16978}\ } (\bibinfo {year} {2024})}\BibitemShut {NoStop}%
\bibitem [{\citenamefont {Ma}\ and\ \citenamefont {Turzillo}(2024)}]{Turzillo_2024_spt}%
  \BibitemOpen
  \bibfield  {author} {\bibinfo {author} {\bibfnamefont {R.}~\bibnamefont {Ma}}\ and\ \bibinfo {author} {\bibfnamefont {A.}~\bibnamefont {Turzillo}},\ }\bibfield  {title} {\bibinfo {title} {Symmetry protected topological phases of mixed states in the doubled space},\ }\href@noop {} {\bibfield  {journal} {\bibinfo  {journal} {arXiv preprint arXiv:2403.13280}\ } (\bibinfo {year} {2024})}\BibitemShut {NoStop}%
\bibitem [{\citenamefont {Xue}\ \emph {et~al.}(2024)\citenamefont {Xue}, \citenamefont {Lee},\ and\ \citenamefont {Bao}}]{Yimu_2024_tn}%
  \BibitemOpen
  \bibfield  {author} {\bibinfo {author} {\bibfnamefont {H.}~\bibnamefont {Xue}}, \bibinfo {author} {\bibfnamefont {J.~Y.}\ \bibnamefont {Lee}},\ and\ \bibinfo {author} {\bibfnamefont {Y.}~\bibnamefont {Bao}},\ }\bibfield  {title} {\bibinfo {title} {Tensor network formulation of symmetry protected topological phases in mixed states},\ }\href@noop {} {\bibfield  {journal} {\bibinfo  {journal} {arXiv preprint arXiv:2403.17069}\ } (\bibinfo {year} {2024})}\BibitemShut {NoStop}%
\bibitem [{\citenamefont {Sohal}\ and\ \citenamefont {Prem}(2024)}]{sohal2024noisy}%
  \BibitemOpen
  \bibfield  {author} {\bibinfo {author} {\bibfnamefont {R.}~\bibnamefont {Sohal}}\ and\ \bibinfo {author} {\bibfnamefont {A.}~\bibnamefont {Prem}},\ }\bibfield  {title} {\bibinfo {title} {A noisy approach to intrinsically mixed-state topological order},\ }\href@noop {} {\bibfield  {journal} {\bibinfo  {journal} {arXiv preprint arXiv:2403.13879}\ } (\bibinfo {year} {2024})}\BibitemShut {NoStop}%
\bibitem [{\citenamefont {Wang}\ and\ \citenamefont {Li}(2024)}]{wang2024anomaly}%
  \BibitemOpen
  \bibfield  {author} {\bibinfo {author} {\bibfnamefont {Z.}~\bibnamefont {Wang}}\ and\ \bibinfo {author} {\bibfnamefont {L.}~\bibnamefont {Li}},\ }\bibfield  {title} {\bibinfo {title} {Anomaly in open quantum systems and its implications on mixed-state quantum phases},\ }\href@noop {} {\bibfield  {journal} {\bibinfo  {journal} {arXiv preprint arXiv:2403.14533}\ } (\bibinfo {year} {2024})}\BibitemShut {NoStop}%
\bibitem [{\citenamefont {Peres}(1996)}]{peres1996}%
  \BibitemOpen
  \bibfield  {author} {\bibinfo {author} {\bibfnamefont {A.}~\bibnamefont {Peres}},\ }\bibfield  {title} {\bibinfo {title} {Separability criterion for density matrices},\ }\href {https://doi.org/10.1103/PhysRevLett.77.1413} {\bibfield  {journal} {\bibinfo  {journal} {Phys. Rev. Lett.}\ }\textbf {\bibinfo {volume} {77}},\ \bibinfo {pages} {1413} (\bibinfo {year} {1996})}\BibitemShut {NoStop}%
\bibitem [{\citenamefont {Horodecki}\ \emph {et~al.}(1996)\citenamefont {Horodecki}, \citenamefont {Horodecki},\ and\ \citenamefont {Horodecki}}]{horodecki1996}%
  \BibitemOpen
  \bibfield  {author} {\bibinfo {author} {\bibfnamefont {M.}~\bibnamefont {Horodecki}}, \bibinfo {author} {\bibfnamefont {P.}~\bibnamefont {Horodecki}},\ and\ \bibinfo {author} {\bibfnamefont {R.}~\bibnamefont {Horodecki}},\ }\bibfield  {title} {\bibinfo {title} {Separability of mixed states: necessary and sufficient conditions},\ }\href {https://doi.org/https://doi.org/10.1016/S0375-9601(96)00706-2} {\bibfield  {journal} {\bibinfo  {journal} {Physics Letters A}\ }\textbf {\bibinfo {volume} {223}},\ \bibinfo {pages} {1 } (\bibinfo {year} {1996})}\BibitemShut {NoStop}%
\bibitem [{\citenamefont {Eisert}\ and\ \citenamefont {Plenio}(1999)}]{eisert99}%
  \BibitemOpen
  \bibfield  {author} {\bibinfo {author} {\bibfnamefont {J.}~\bibnamefont {Eisert}}\ and\ \bibinfo {author} {\bibfnamefont {M.~B.}\ \bibnamefont {Plenio}},\ }\bibfield  {title} {\bibinfo {title} {A comparison of entanglement measures},\ }\href {https://doi.org/10.1080/09500349908231260} {\bibfield  {journal} {\bibinfo  {journal} {Journal of Modern Optics}\ }\textbf {\bibinfo {volume} {46}},\ \bibinfo {pages} {145} (\bibinfo {year} {1999})}\BibitemShut {NoStop}%
\bibitem [{\citenamefont {Vidal}\ and\ \citenamefont {Werner}(2002)}]{vidal2002}%
  \BibitemOpen
  \bibfield  {author} {\bibinfo {author} {\bibfnamefont {G.}~\bibnamefont {Vidal}}\ and\ \bibinfo {author} {\bibfnamefont {R.~F.}\ \bibnamefont {Werner}},\ }\bibfield  {title} {\bibinfo {title} {Computable measure of entanglement},\ }\href {https://doi.org/10.1103/PhysRevA.65.032314} {\bibfield  {journal} {\bibinfo  {journal} {Phys. Rev. A}\ }\textbf {\bibinfo {volume} {65}},\ \bibinfo {pages} {032314} (\bibinfo {year} {2002})}\BibitemShut {NoStop}%
\bibitem [{\citenamefont {Lu}\ and\ \citenamefont {Vijay}(2023)}]{lu2022_lre}%
  \BibitemOpen
  \bibfield  {author} {\bibinfo {author} {\bibfnamefont {T.-C.}\ \bibnamefont {Lu}}\ and\ \bibinfo {author} {\bibfnamefont {S.}~\bibnamefont {Vijay}},\ }\bibfield  {title} {\bibinfo {title} {Characterizing long-range entanglement in a mixed state through an emergent order on the entangling surface},\ }\href {https://doi.org/10.1103/PhysRevResearch.5.033031} {\bibfield  {journal} {\bibinfo  {journal} {Phys. Rev. Res.}\ }\textbf {\bibinfo {volume} {5}},\ \bibinfo {pages} {033031} (\bibinfo {year} {2023})}\BibitemShut {NoStop}%
\bibitem [{\citenamefont {Chen}\ \emph {et~al.}(2011{\natexlab{a}})\citenamefont {Chen}, \citenamefont {Gu},\ and\ \citenamefont {Wen}}]{spt_1d_2011}%
  \BibitemOpen
  \bibfield  {author} {\bibinfo {author} {\bibfnamefont {X.}~\bibnamefont {Chen}}, \bibinfo {author} {\bibfnamefont {Z.-C.}\ \bibnamefont {Gu}},\ and\ \bibinfo {author} {\bibfnamefont {X.-G.}\ \bibnamefont {Wen}},\ }\bibfield  {title} {\bibinfo {title} {Classification of gapped symmetric phases in one-dimensional spin systems},\ }\href {https://doi.org/10.1103/PhysRevB.83.035107} {\bibfield  {journal} {\bibinfo  {journal} {Phys. Rev. B}\ }\textbf {\bibinfo {volume} {83}},\ \bibinfo {pages} {035107} (\bibinfo {year} {2011}{\natexlab{a}})}\BibitemShut {NoStop}%
\bibitem [{\citenamefont {Chen}\ \emph {et~al.}(2011{\natexlab{b}})\citenamefont {Chen}, \citenamefont {Gu},\ and\ \citenamefont {Wen}}]{spt_2011}%
  \BibitemOpen
  \bibfield  {author} {\bibinfo {author} {\bibfnamefont {X.}~\bibnamefont {Chen}}, \bibinfo {author} {\bibfnamefont {Z.-C.}\ \bibnamefont {Gu}},\ and\ \bibinfo {author} {\bibfnamefont {X.-G.}\ \bibnamefont {Wen}},\ }\bibfield  {title} {\bibinfo {title} {Complete classification of one-dimensional gapped quantum phases in interacting spin systems},\ }\href {https://doi.org/10.1103/PhysRevB.84.235128} {\bibfield  {journal} {\bibinfo  {journal} {Phys. Rev. B}\ }\textbf {\bibinfo {volume} {84}},\ \bibinfo {pages} {235128} (\bibinfo {year} {2011}{\natexlab{b}})}\BibitemShut {NoStop}%
\bibitem [{\citenamefont {Briegel}\ and\ \citenamefont {Raussendorf}(2001)}]{Raussendorf_2001_ghz}%
  \BibitemOpen
  \bibfield  {author} {\bibinfo {author} {\bibfnamefont {H.~J.}\ \bibnamefont {Briegel}}\ and\ \bibinfo {author} {\bibfnamefont {R.}~\bibnamefont {Raussendorf}},\ }\bibfield  {title} {\bibinfo {title} {Persistent entanglement in arrays of interacting particles},\ }\href {https://doi.org/10.1103/PhysRevLett.86.910} {\bibfield  {journal} {\bibinfo  {journal} {Phys. Rev. Lett.}\ }\textbf {\bibinfo {volume} {86}},\ \bibinfo {pages} {910} (\bibinfo {year} {2001})}\BibitemShut {NoStop}%
\bibitem [{\citenamefont {Lu}\ and\ \citenamefont {Grover}(2020)}]{lu2019structure}%
  \BibitemOpen
  \bibfield  {author} {\bibinfo {author} {\bibfnamefont {T.-C.}\ \bibnamefont {Lu}}\ and\ \bibinfo {author} {\bibfnamefont {T.}~\bibnamefont {Grover}},\ }\bibfield  {title} {\bibinfo {title} {Structure of quantum entanglement at a finite temperature critical point},\ }\href {https://doi.org/10.1103/PhysRevResearch.2.043345} {\bibfield  {journal} {\bibinfo  {journal} {Phys. Rev. Research}\ }\textbf {\bibinfo {volume} {2}},\ \bibinfo {pages} {043345} (\bibinfo {year} {2020})}\BibitemShut {NoStop}%
\bibitem [{\citenamefont {Lu}\ \emph {et~al.}(2020)\citenamefont {Lu}, \citenamefont {Hsieh},\ and\ \citenamefont {Grover}}]{Lu_topo_nega_2020}%
  \BibitemOpen
  \bibfield  {author} {\bibinfo {author} {\bibfnamefont {T.-C.}\ \bibnamefont {Lu}}, \bibinfo {author} {\bibfnamefont {T.~H.}\ \bibnamefont {Hsieh}},\ and\ \bibinfo {author} {\bibfnamefont {T.}~\bibnamefont {Grover}},\ }\bibfield  {title} {\bibinfo {title} {Detecting topological order at finite temperature using entanglement negativity},\ }\href {https://doi.org/10.1103/PhysRevLett.125.116801} {\bibfield  {journal} {\bibinfo  {journal} {Phys. Rev. Lett.}\ }\textbf {\bibinfo {volume} {125}},\ \bibinfo {pages} {116801} (\bibinfo {year} {2020})}\BibitemShut {NoStop}%
\bibitem [{\citenamefont {Kitaev}\ and\ \citenamefont {Preskill}(2006)}]{Kitaev06_1}%
  \BibitemOpen
  \bibfield  {author} {\bibinfo {author} {\bibfnamefont {A.}~\bibnamefont {Kitaev}}\ and\ \bibinfo {author} {\bibfnamefont {J.}~\bibnamefont {Preskill}},\ }\bibfield  {title} {\bibinfo {title} {Topological entanglement entropy},\ }\href {https://doi.org/10.1103/PhysRevLett.96.110404} {\bibfield  {journal} {\bibinfo  {journal} {Phys. Rev. Lett.}\ }\textbf {\bibinfo {volume} {96}},\ \bibinfo {pages} {110404} (\bibinfo {year} {2006})}\BibitemShut {NoStop}%
\bibitem [{\citenamefont {You}\ \emph {et~al.}(2014)\citenamefont {You}, \citenamefont {Bi}, \citenamefont {Rasmussen}, \citenamefont {Slagle},\ and\ \citenamefont {Xu}}]{You_strange_correlator_2014}%
  \BibitemOpen
  \bibfield  {author} {\bibinfo {author} {\bibfnamefont {Y.-Z.}\ \bibnamefont {You}}, \bibinfo {author} {\bibfnamefont {Z.}~\bibnamefont {Bi}}, \bibinfo {author} {\bibfnamefont {A.}~\bibnamefont {Rasmussen}}, \bibinfo {author} {\bibfnamefont {K.}~\bibnamefont {Slagle}},\ and\ \bibinfo {author} {\bibfnamefont {C.}~\bibnamefont {Xu}},\ }\bibfield  {title} {\bibinfo {title} {Wave function and strange correlator of short-range entangled states},\ }\href {https://doi.org/10.1103/PhysRevLett.112.247202} {\bibfield  {journal} {\bibinfo  {journal} {Phys. Rev. Lett.}\ }\textbf {\bibinfo {volume} {112}},\ \bibinfo {pages} {247202} (\bibinfo {year} {2014})}\BibitemShut {NoStop}%
\bibitem [{\citenamefont {Pollmann}\ and\ \citenamefont {Turner}(2012)}]{Pollmann_2012_string_order}%
  \BibitemOpen
  \bibfield  {author} {\bibinfo {author} {\bibfnamefont {F.}~\bibnamefont {Pollmann}}\ and\ \bibinfo {author} {\bibfnamefont {A.~M.}\ \bibnamefont {Turner}},\ }\bibfield  {title} {\bibinfo {title} {Detection of symmetry-protected topological phases in one dimension},\ }\href {https://doi.org/10.1103/PhysRevB.86.125441} {\bibfield  {journal} {\bibinfo  {journal} {Phys. Rev. B}\ }\textbf {\bibinfo {volume} {86}},\ \bibinfo {pages} {125441} (\bibinfo {year} {2012})}\BibitemShut {NoStop}%
\bibitem [{\citenamefont {Onsager}(1944)}]{onsager_1944}%
  \BibitemOpen
  \bibfield  {author} {\bibinfo {author} {\bibfnamefont {L.}~\bibnamefont {Onsager}},\ }\bibfield  {title} {\bibinfo {title} {Crystal statistics. i. a two-dimensional model with an order-disorder transition},\ }\href {https://doi.org/10.1103/PhysRev.65.117} {\bibfield  {journal} {\bibinfo  {journal} {Phys. Rev.}\ }\textbf {\bibinfo {volume} {65}},\ \bibinfo {pages} {117} (\bibinfo {year} {1944})}\BibitemShut {NoStop}%
\bibitem [{\citenamefont {Raussendorf}\ \emph {et~al.}(2005)\citenamefont {Raussendorf}, \citenamefont {Bravyi},\ and\ \citenamefont {Harrington}}]{3d_cluster_state_2005}%
  \BibitemOpen
  \bibfield  {author} {\bibinfo {author} {\bibfnamefont {R.}~\bibnamefont {Raussendorf}}, \bibinfo {author} {\bibfnamefont {S.}~\bibnamefont {Bravyi}},\ and\ \bibinfo {author} {\bibfnamefont {J.}~\bibnamefont {Harrington}},\ }\bibfield  {title} {\bibinfo {title} {Long-range quantum entanglement in noisy cluster states},\ }\href {https://doi.org/10.1103/PhysRevA.71.062313} {\bibfield  {journal} {\bibinfo  {journal} {Phys. Rev. A}\ }\textbf {\bibinfo {volume} {71}},\ \bibinfo {pages} {062313} (\bibinfo {year} {2005})}\BibitemShut {NoStop}%
\bibitem [{\citenamefont {Wegner}(1971)}]{wegner_1971}%
  \BibitemOpen
  \bibfield  {author} {\bibinfo {author} {\bibfnamefont {F.~J.}\ \bibnamefont {Wegner}},\ }\bibfield  {title} {\bibinfo {title} {{Duality in Generalized Ising Models and Phase Transitions without Local Order Parameters}},\ }\href {https://doi.org/10.1063/1.1665530} {\bibfield  {journal} {\bibinfo  {journal} {Journal of Mathematical Physics}\ }\textbf {\bibinfo {volume} {12}},\ \bibinfo {pages} {2259} (\bibinfo {year} {1971})}\BibitemShut {NoStop}%
\bibitem [{\citenamefont {Kogut}(1979)}]{1979_kogut}%
  \BibitemOpen
  \bibfield  {author} {\bibinfo {author} {\bibfnamefont {J.~B.}\ \bibnamefont {Kogut}},\ }\bibfield  {title} {\bibinfo {title} {An introduction to lattice gauge theory and spin systems},\ }\href {https://doi.org/10.1103/RevModPhys.51.659} {\bibfield  {journal} {\bibinfo  {journal} {Rev. Mod. Phys.}\ }\textbf {\bibinfo {volume} {51}},\ \bibinfo {pages} {659} (\bibinfo {year} {1979})}\BibitemShut {NoStop}%
\bibitem [{\citenamefont {Vijay}\ \emph {et~al.}(2016)\citenamefont {Vijay}, \citenamefont {Haah},\ and\ \citenamefont {Fu}}]{xcube}%
  \BibitemOpen
  \bibfield  {author} {\bibinfo {author} {\bibfnamefont {S.}~\bibnamefont {Vijay}}, \bibinfo {author} {\bibfnamefont {J.}~\bibnamefont {Haah}},\ and\ \bibinfo {author} {\bibfnamefont {L.}~\bibnamefont {Fu}},\ }\bibfield  {title} {\bibinfo {title} {Fracton topological order, generalized lattice gauge theory, and duality},\ }\href {https://doi.org/10.1103/PhysRevB.94.235157} {\bibfield  {journal} {\bibinfo  {journal} {Phys. Rev. B}\ }\textbf {\bibinfo {volume} {94}},\ \bibinfo {pages} {235157} (\bibinfo {year} {2016})}\BibitemShut {NoStop}%
\bibitem [{\citenamefont {Haah}(2011)}]{Haah}%
  \BibitemOpen
  \bibfield  {author} {\bibinfo {author} {\bibfnamefont {J.}~\bibnamefont {Haah}},\ }\bibfield  {title} {\bibinfo {title} {Local stabilizer codes in three dimensions without string logical operators},\ }\href {https://doi.org/10.1103/PhysRevA.83.042330} {\bibfield  {journal} {\bibinfo  {journal} {Phys. Rev. A}\ }\textbf {\bibinfo {volume} {83}},\ \bibinfo {pages} {042330} (\bibinfo {year} {2011})}\BibitemShut {NoStop}%
\bibitem [{\citenamefont {Hsieh}(2016)}]{hsieh_2012_mapping}%
  \BibitemOpen
  \bibfield  {author} {\bibinfo {author} {\bibfnamefont {T.~H.}\ \bibnamefont {Hsieh}},\ }\bibfield  {title} {\bibinfo {title} {From d-dimensional quantum to d+ 1-dimensional classical systems},\ }\href {https://www.semanticscholar.org/paper/From-d-dimensional-Quantum-to-d%2B-1-dimensional-Hsieh/5647185023db3e7695a1309bbf39ee2f0b0a7fb8} {\bibfield  {journal} {\bibinfo  {journal} {Student Review}\ } (\bibinfo {year} {2016})}\BibitemShut {NoStop}%
\end{thebibliography}%
	
	\newpage 
	\appendix

\onecolumngrid

\begin{center}
\textbf{APPENDIX}
\end{center}
In Appendix.\ref{append:spectrum}, we present a unified derivation of negativity spectrum and negativity for $d$-dimensional toric codes ($d=2,3,4$) subject to boundary decoherence. Appendix.\ref{append:SPT} discusses the stability of the perturbed SPTs encoded in the negativity spectrum. Appendix.\ref{append:qc_mapping} presents the mapping between a 2d quantum Ising gauge theory at zero temperature and a 3d classical Ising gauge theory at finite temperature.  Appendix.\ref{append:xz_both} presents the calculation of the negativity spectrum of the 2d toric code subject to both Pauli-X and Pauli-Z decoherence on the boundary.

\section{General formalism of computing negativity spectrum}\label{append:spectrum}

Here we present the calculation of the negativity spectrum for the toric code in $d$ space dimensions subject to the decoherence acting on the $d-1$ dimensional bipartition boundary that separates the subregions $\mA$ and $\mB$.

The toric code Hamiltonian in $d$ space dimensions takes the form $H=- \sum_i A_i - \sum_j B_j$, where the stabilizer $A_i$ is the product of Pauli-Zs around $i$, and the stabilizer $B_j$ is the product of Pauli-Xs around $j$:
\vspace{2mm}

\textbf{2d toric code}: qubits are defined on 1-cells, $A_i$ is a product of Pauli-Zs on four 1-cells emanating from a 0-cell, and $B_j$ is a product of Pauli-Xs on four 1-cells on the boundary of a 2-cell.

\textbf{3d toric code}: qubits are defined on 1-cells, $A_i$ is a product of Pauli-Zs on six 1-cells emanating from a 0-cell, and $B_j$ is a product of Pauli-Xs on four 1-cells on the boundary of a 2-cell.

\textbf{4d toric code}: qubits are defined on 2-cells, $A_i$ is a product of Pauli-Zs on the six 2-cells adjacent to a 1-cell, and $B_j$ is the product of Pauli-Xs on the six 2-cells on the boundary of a 3-cell. 

\vspace{2mm}

The corresponding ground-state subspace density matrix is $\rho_0  \propto\prod_{i} \frac{1+A_i}{2} \prod_j  \frac{1+B_j}{2} \propto  \sum_{a,b}  \prod_i A_i^{a_i } \prod_j B_j^{b_j }$, with $a\equiv\{ a_i \}, b= \{b_j \}$, and $a_i, b_j= 0,1$.

\subsection{Pauli-Z decoherence}\label{append:sec_z}
We consider the Pauli-Z noise channels applied on the set of qubits (denoted by $\mathcal{M}$) that only excites the $B_j$ stabilizers on the boundary. At error rate $p_z$, the decohered mixed state $\rho= \prod_{i\in M} N_{z,i}[\rho_0]$, with $N_{z,i}[\rho_0] = (1-p_z) \rho_0 + p_z Z_i \rho_0 Z_i$, may be expressed as

\begin{equation}
\rho \propto \sum_{a,b}  \prod_i  A_i^{a_i } \prod_j  B_j^{b_j } (1-2p)^{W(X)}.
\end{equation}
The stabilizer string $\prod_j  B_j^{b_j } $ may be factorized into a product of the bulk part $ \prod_j^{\text{bulk}}  B_j^{b_j }$ and the boundary part $ \prod_j^{\partial}  B_j^{b_j } $, and $W(X)$ is the number of Pauli-Xs belonging to the set $\mathcal{M}$ that appear in the boundary stabilizer string $\prod_j^{\partial}  B_j^{b_j }$. 

Taking a partial transpose (e.g. a transpose acting on the subregion $\mA$) gives the matrix 

\begin{equation}
\rho^{\Gamma} \propto \sum_{a,b}  \prod_i  A_i^{a_i } \prod_j  B_j^{b_j } (1-2p)^{W(X)} \psi(a,b), 
\end{equation}
with $\psi(a,b) = 1/-1 $ when the stabilizer strings $\prod_i  A_i^{a_i } \prod_j  B_j^{b_j }$ has even/odd number of Pauli-Ys in the subregion $\mA$. Since both $W(X)$ and $\psi(a,b)$ only depends on $a_i,b_j$ of the boundary stabilizers, one can sum over $a_i, b_j$ in the bulk, resulting in

\begin{equation}
\rho^{\Gamma} \propto   \left(\prod_{i \in \text{bulk}} \frac{1+A_i}{2} \prod_{j \in \text{bulk}}  \frac{1+B_j}{2} \right)  \left( \sum_{a,b}^{\partial}   \prod_i  A_i^{a_i } \prod_j  B_j^{b_j } (1-2p)^{W(X)} \psi(a,b)\right) 
\end{equation}

The negativity spectrum, i.e. the eigenvalues of $\rho^{\Gamma}$, can be obtained by replacing all stabilizers by $1$ or $-1$.  As the stabilizers in the bulk are fixed at $1$, the negativity spectrum is solely determined by the values of the boundary stabilizers. Therefore we write the spectrum as

\begin{equation}
\rho^{\Gamma} \propto  \sum_{a,b}^{\partial}   \prod_i  A_i^{a_i } \prod_j  B_j^{b_j } (1-2p)^{W(X)} \psi(a,b) ,  
\end{equation}
where we abuse the notation by treating $\rho^{\Gamma}$ as the spectrum, rather than a matrix.

To compute $W(X)$ and $\psi(a,b)$, we  locate the stabilizers on the $d-1$ dimensional bipartition boundary as follows:

\begin{equation*}
\includegraphics[width=10cm]{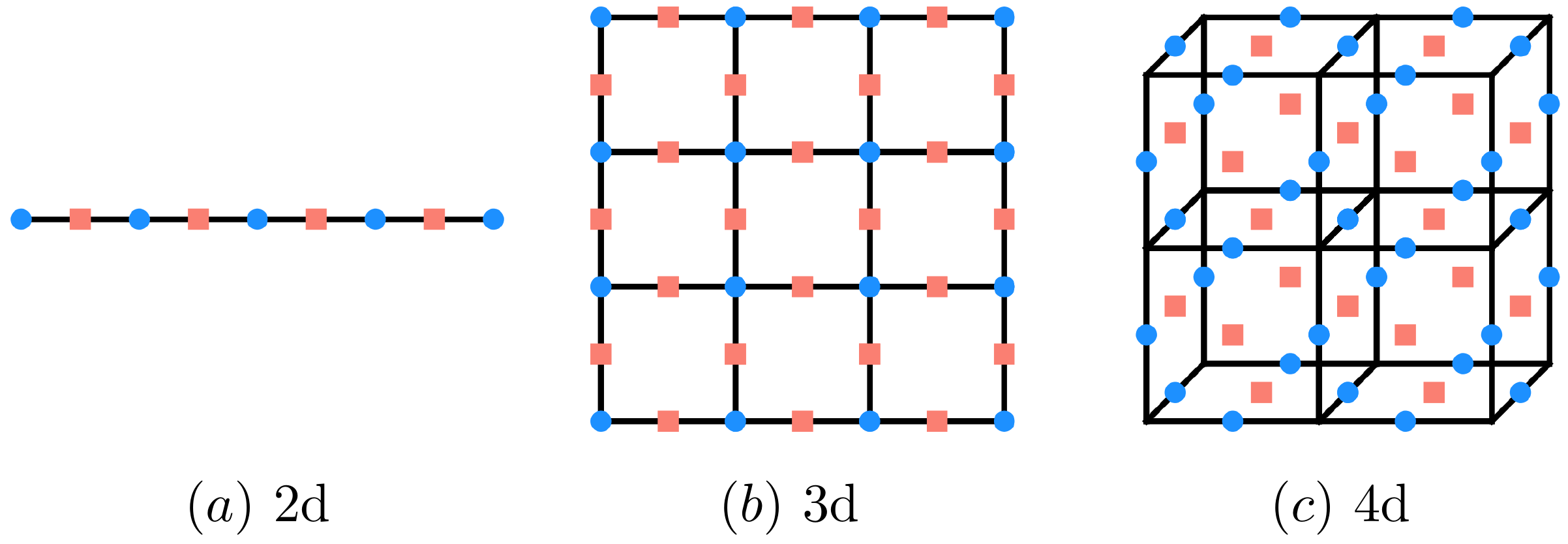}
\end{equation*} 
where (a), (b), and (c) depict the boundary of the toric code in 2d, 3d, and 4d, respectively; the set of blue circles (denoted as $R_A$) specifies the boundary $A_i$ stabilizers, and the set of red squares (denoted as $R_B$) specifies the boundary $B_j$ stabilizers. 

As mentioned above, $W(X)$ counts the number of Pauli-Xs in the set $\mathcal{M}$ appearing in stabilizer string $\prod_j^{\partial}  B_j^{b_j }$. This gives $W(X)=  \frac{1}{2}\sum_{i\in R_A} (1- \prod_{j \in \partial^* i  } \tau_j   )$, where $\tau_j \equiv (-1)^{b_j}$, and $\prod_{j \in \partial^* i  } $ denotes the product over $j (\in R_B)$ on the coboundary of $i \in R_A$ (i.e. $i$ belongs to the boundary of $j$  ). For example, for 2d boundary of the 3d toric code, $W(X) = \frac{1}{2} \sum_{i\in R_A}  (1- \tau \tau\tau\tau  )  $, where $\tau \tau\tau\tau$ denotes the product of $\tau_j$s on the four edges emanating from the vertex (0-cell).

On the other hand, $\psi(a,b) = (-1)^{N_Y}$, where $N_Y$ is the number of Pauli-Ys in the subregion $\mA$ in the stabilizer string $\prod_{  i \in R_A } A_i^{a_i}   \prod_{  j \in R_B } B_j^{b_j}$. Since the simultaneous presence of neighboring $A_i$, $B_j$ stabilizers contributes an odd number of Pauli-Ys, one has $\psi(a,b) =\prod_{ \expval{i\in R_A,j \in R_B }  }  (-1)^{a_ib_j}   $, where $\expval{i\in R_A,j \in R_B }$ denotes the pair $i\in R_A, j\in R_B$ neighboring to each other. 
 
With the results of $W(X), \psi(a,b)$,  we derive the negativity spectrum:
\begin{equation}
\rho^{\Gamma} \propto  \sum_{a,b}   \prod_{i\in R_A}  A_i^{a_i } \prod_{j\in R_B}  B_j^{b_j } e^{ \beta_z  
 \sum_{i \in R_A} \prod_{j\in \partial^* i} \tau_j    }\psi(a,b),  
\end{equation}
with $\beta_z = - \frac{1}{2} \log(1-2 p_z)$. 

The structure of the negativity spectrum may be understood by introducing the fictitious Hilbert space of qubits living on the sublattices $R_A, R_B$ spanned by the Z basis product states $\ket{a,b} =  \ket{\{a_i,b_j \}}$. We can define a state $\ket{\psi} \propto \sum_{a,b} \psi(a,b)\ket{a,b}$ and a trivial state $\ket{+} \propto \sum_{a,b} \ket{a,b}$, then the negativity spectrum can be written as 

\begin{equation}\label{appendix:z_noise_spectrum}
\rho^{\Gamma}  \propto \bra{+ } \prod_{i\in R_A}Z_{A,i}^{\frac{1-A_i}{2} } \prod_{j\in R_B}Z_{B,j}^{\frac{1-B_j}{2} } e^{\beta_z \sum_{i\in R_A} \prod_{j\in \partial^* i } Z_{B,j}   } \ket{\psi},
\end{equation}
where distinct eigenvalues in the negativity spectrum correspond to different patterns of inserting $Z_{i,A}, Z_{j,B}$ acting on the fictitious Hilbert space. 

Since $\ket{\psi}= \sum_{a,b} \prod_{\expval{ i \in R_A, j \in R_B}  }  (-1)^{  a_i b_j  }   \ket{a,b}$, $\ket{\psi}$ can be obtained by applying $\prod_{\expval{ i \in R_A, j\in R_B}} \text{CZ}_{i,j}$ on the trivial state $ \ket{+ }$, with $\text{CZ}_{i,j}$ being the two-qubit controlled-Z gates. It follows that $\ket{\psi}$ is the ground state of the cluster-state Hamiltonian $H_c=- \sum_{i\in R_A} X_{A,i} \prod_{j \in \partial^* i} Z_{B,j} - \sum_{j\in R_B} X_{B,j} \prod_{i \in \partial j} Z_{A,i}$. Therefore, one may employ the  condition 
$ \prod_{j\in \partial^* i } Z_{B,j} \ket{\psi}  = X_{A,i} \ket{\psi }$ to simplify the negativity spectrum: 

\begin{equation}
\begin{split}
\rho^{\Gamma}  &\propto \bra{+ } \prod_{i\in R_A}Z_{A,i}^{\frac{1-A_i}{2} } \prod_{j\in R_B}Z_{B,j}^{\frac{1-B_j}{2} } e^{\beta_z \sum_{i\in R_A}  X_{A,i}   } \ket{\psi} \\
& = \bra{+}\prod_{i\in R_A}Z_{A,i}^{\frac{1-A_i}{2} } \prod_{j\in R_B}Z_{B,j}^{\frac{1-B_j}{2} } \ket{\psi} e^{\beta_z  \sum_{i\in R_A} A_i   }.  
\end{split}
\end{equation}

Note that to have non-vanishing eigenvalues, the configurations $A \equiv \{A_i\}, B\equiv \{ B_j \}$ need to obey the symmetry constraint resulting from the symmetry of the cluster-state Hamiltonian. Specifically, $\ket{\psi}$ has two types of symmetries: $\prod_{i \in  \Gamma^*_A } X_{A,i}=1$ and $ \prod_{j \in \Gamma_B} X_{B,j}  =1$, where $\Gamma_A^*$ is a cocycle (which has trivial coboundary, i.e. $\partial^* \Gamma_A^*=0$), and $\Gamma_B$ is a cycle (which has trivial boundary, i.e. $\partial \Gamma_B=0$). Hence the configurations $A,B$ must satisfy $\prod_{i \in  \Gamma^*_A }A_i =1$ and $ \prod_{j \in \Gamma_B}B_j=1$, to have non-vanishing eigenvalues in the negativity spectrum.  

Finally, the negativity spectrum reads 

\begin{equation}
\rho^{\Gamma} =  \frac{1}{Z}  \chi(A,B)  e^{\beta_z  \sum_{i\in R_A} A_i   }, 
\end{equation}
where $Z = \sum_{A,B} \chi(A,B)  e^{\beta_z  \sum_{i\in R_A} A_i   }   $ is a normalization constant to ensure that the negativity spectrum sums to one, and $\chi(A,B) = \frac{ \bra{+}\prod_{i\in R_A}Z_{A,i}^{\frac{1-A_i}{2} } \prod_{j\in R_B}Z_{B,j}^{\frac{1-B_j}{2} }  \ket{\psi}}{ \bra{+}  \ket{\psi}   }  = 1,-1$ for $\{A_i, B_j\}$ configurations that satisfy $\prod_{i \in  \Gamma^*_A }A_i =1$ and $ \prod_{j \in \Gamma_B}B_j=1$.

With the negativity spectrum, we proceed to compute entanglement negativity $E_N= \log ( \abs{\rho^{\Gamma}}_1  )$, with 

\begin{equation}
\abs{\rho^{\Gamma}}_1= \frac{1}{Z} \sum_{A,B}\abs{\chi(A,B)  e^{\beta_z  \sum_{i\in R_A} A_i   }}  
\end{equation}

First, we compute the denominator $Z$:

\begin{equation}
\begin{split}
Z&=  \sum_{A,B}  \frac{ \bra{+}\prod_{i\in R_A}Z_{A,i}^{\frac{1-A_i}{2} } \prod_{j\in R_B}Z_{B,j}^{\frac{1-B_j}{2} } e^{\beta_z \sum_{i\in R_A } \prod_{j \in \partial^* i} Z_{B,j}  }\ket{\psi}}{ \bra{+}  \ket{\psi}   } \\ 
& = \frac{1}{ \bra{+}  \ket{\psi}  }\bra{+}\prod_{i\in R_A}(1+Z_{A,i}) \prod_{j\in R_B} (1+Z_{B,j} ) e^{\beta_z \sum_{i\in R_A } \prod_{j \in \partial^* i} Z_{B,j}  }\ket{\psi}
\end{split}
\end{equation}
Since $\ket{\psi} =  \prod_{\expval{ i \in R_A, j\in R_B}} \text{CZ}_{i,j} \ket{+ }$, one has $ \bra{+}\prod_{i\in R_A}  ( 1+ Z_{A,i}) \prod_{j\in R_B} (1+Z_{B,j})    \ket{\psi} =\bra{+}\prod_{i\in R_A}  ( 1+ Z_{A,i}) \prod_{j\in R_B} (1+Z_{B,j})    \ket{+}  = 1 $. As a result,

\begin{equation}
Z= \frac{1}{ \bra{+} \ket{\psi}} e^{\beta_z \sum_{i\in R_A }} .
\end{equation}

On the other hand, $\sum_{A,B}\abs{\chi(A,B)  e^{\beta_z  \sum_{i\in R_A} A_i}} =\sum_{A,B}'  e^{\beta_z  \sum_{i\in R_A} A_i}$, where $\sum_{A,B}'$ denotes the summation over the $A,B$ configurations with non-zero $\chi(A,B)$. Combining with the calculation of $Z$, one has

\begin{equation}
\abs{\rho^{\Gamma}}_1= \bra{+} \ket{\psi}
\frac{\sum_A' \sum_B' e^{\beta_z  \sum_{i\in R_A} A_i   }}{e^{ \beta_z \sum_{i\in R_A} } }
\end{equation}
This can be further simplified by noticing that $ \bra{+} \ket{\psi} \sum_B' = \sum_B  \bra{ +} \prod_{j\in R_B }Z_{B,j}^{\frac{1-B_j}{2}} \ket{\psi}  =  \bra{ +} \prod_{j\in R_B } (1+Z_{B,j}) \prod_{\expval{i \in R_A, j \in R_B}}\text{CZ}_{ij} \ket{+}=  \bra{ +} \prod_{j\in R_B } (1+Z_{B,j})\ket{+} =1$. Consequently,  

\begin{equation}
\abs{\rho^{\Gamma}}_1= 
\frac{\sum_A'  e^{\beta_z  \sum_{i\in R_A} A_i   }}{  \sum_{A}' e^{ \beta_z \sum_{i\in R_A} A_i} \prod_{i\in R_A} \delta(A_i=1 ) }
\end{equation}

As discussed above, $\{A\}$ needs to obey the constraint $\prod_{i \in  \Gamma^*_A }A_i=1$, and this can be resolved by defining $A_i = \prod_{j \in \partial^* i } \tau_j $, where $\{\tau_j\}$ are independent Ising variables defined on $R_B$. As a result, $\abs{\rho^{\Gamma}}_1=  \frac{Z_B}{\tilde{Z}_B}$

\begin{equation}
\begin{split}
&Z_B= \sum_\tau  e^{\beta_z  \sum_{i\in R_A} \prod_{j \in  \partial^* i } \tau_j }\\
&\tilde{Z}_B = \sum_{\tau} e^{ \beta_z \sum_{i\in R_A } \prod_{j \in \partial^* i}\tau_j      } \prod_{i\in R_A} \delta(\prod_{j \in \partial^* i}\tau_j =1)  
\end{split}
\end{equation} 
Hence the entanglement negativity is the free energy difference associated with forbidding excitations in a certain classical statistical mechanics model:

\begin{equation}\label{append:negativity_z}
E_N = \log Z_B -  \log \tilde{Z}_B
\end{equation}

\subsection{Pauli-X decoherence}
Here we consider the Pauli-X noise applied on the set of qubits (denoted by $\mathcal{K}$) that only excites the $A_i$ stabilizers on the boundary. At error rate $p_x$, the decohered mixed state $\rho= \prod_{i\in \mathcal{K} } N_{x,i}[\rho_0]$, with $N_{x,i}[\rho_0] = (1-p_x) \rho_0 + p_x X_i \rho_0 X_i$, may be expressed as

\begin{equation}
\rho \propto \sum_{a,b}  \prod_i  A_i^{a_i } \prod_j  B_j^{b_j } (1-2p)^{W(Z)}.
\end{equation}
$W(Z)$ is the number of Pauli-Zs belonging to the set $\mathcal{K}$ that appear in the boundary stabilizer string $\prod_i^{\partial}  A_i^{a_i }$, which can be computed as $W(Z) =  \frac{1}{2}\sum_{j \in R_B}  (1- \prod_{i \in \partial j  } \sigma_i  )$, where $\sigma_i  = (-1)^{a_i}$, and  $\prod_{i \in \partial j  } \sigma_i$ denotes the product of $\sigma_i$ on the boundary of $j \in R_B$. 

With a calculation similar to the Pauli-Z decoherence, we derive the negativity spectrum:
\begin{equation}
\rho^{\Gamma} \propto  \sum_{a,b}   \prod_{i\in R_A}  A_i^{a_i } \prod_{j\in R_B}  B_j^{b_j } e^{ \beta_x  
 \sum_{j \in R_B} \prod_{i\in \partial j} \sigma_i    }\psi(a,b),  
\end{equation}
with $\beta_x = - \frac{1}{2} \log(1-2 p_x)$.  

Again, we can introduce the fictitious Hilbert space of qubits living on the lattices $R_A, R_B$ spanned by the Z basis product states $\ket{a,b} =  \ket{\{a_i,b_j \}}$. Defining a state $\ket{\psi} \propto\sum_{a,b} \psi(a,b)\ket{a,b}$ and a trivial state $\ket{+} \propto \sum_{a,b} \ket{a,b}$, the negativity spectrum can be written as 

\begin{equation}\label{append:x_spectrum}
\rho^{\Gamma}  \propto \bra{+ } \prod_{i\in R_A}Z_{A,i}^{\frac{1-A_i}{2} } \prod_{j\in R_B}Z_{B,j}^{\frac{1-B_j}{2} } e^{\beta_x \sum_{j\in R_B} \prod_{i\in \partial j  } Z_{A,i}   } \ket{\psi},
\end{equation}

By employing the  condition 
$ \prod_{i\in \partial j  } Z_{A,i} \ket{\psi}  = X_{B,j} \ket{\psi }$, the negativity spectrum can be simplified as:  

\begin{equation}
\begin{split}
\rho^{\Gamma}  &\propto \bra{+ } \prod_{i\in R_A}Z_{A,i}^{\frac{1-A_i}{2} } \prod_{j\in R_B}Z_{B,j}^{\frac{1-B_j}{2} } e^{\beta_x \sum_{j\in R_B}  X_{B,j}   } \ket{\psi} \\
& = \bra{+}\prod_{i\in R_A}Z_{A,i}^{\frac{1-A_i}{2} } \prod_{j\in R_B}Z_{B,j}^{\frac{1-B_j}{2} } \ket{\psi} e^{\beta_x  \sum_{j\in R_B} B_j   }.  
\end{split}
\end{equation}

Finally, the negativity spectrum reads 
\begin{equation}
\rho^{\Gamma} =  \frac{1}{Z}  \chi(A,B)  e^{\beta_x  \sum_{j\in R_B} B_j  }, 
\end{equation}
where $Z = \sum_{A,B} \chi(A,B)  e^{\beta_x  \sum_{j \in R_B} B_j }   $ is a normalization constant, and $\chi(A,B) = \frac{ \bra{+}\prod_{i\in R_A}Z_{A,i}^{\frac{1-A_i}{2} } \prod_{j\in R_B}Z_{B,j}^{\frac{1-B_j}{2} }  \ket{\psi}}{ \bra{+}  \ket{\psi}   }  = 1,-1$ for $\{A_i, B_j\}$ configurations that satisfy $\prod_{i \in  \Gamma^*_A }A_i =1$ and $ \prod_{j \in \Gamma_B}B_j=1$.

With the negativity spectrum, by following a calculation similar to Appendix.\ref{append:sec_z}, we derive the entanglement negativity

\begin{equation}
\abs{\rho^{\Gamma}}_1= 
\frac{\sum_B'  e^{\beta_x  \sum_{j\in R_B} B_j   }}{  \sum_B' e^{ \beta_x \sum_{j\in R_B} B_j } \prod_{j\in R_B} \delta(B_j=1 ) }
\end{equation}

As discussed above, $\{B_j\}$ needs to obey the constraint $\prod_{j \in  \Gamma_B }B_j=1$, and this can be resolved by defining $B_j = \prod_{i \in \partial j  } \sigma_i$, where $\{\sigma_i \}$ are independent Ising variables defined on $R_A$. As a result, $\abs{\rho^{\Gamma}}_1=  \frac{Z_A}{\tilde{Z}_A}$

\begin{equation}
\begin{split}
&Z_A= \sum_\sigma  e^{\beta_x  \sum_{j \in R_B} \prod_{i \in  \partial j  } \sigma_i }\\
&\tilde{Z}_A = \sum_{\sigma} e^{ \beta_x \sum_{j\in R_B}  \prod_{ i \in \partial j }  \sigma_i  }   \prod_{j \in R_B } \delta(  \prod_{i\in \partial j } \sigma_i =1  ).
\end{split}
\end{equation} 
Hence the entanglement negativity is the free energy difference associated with forbidding excitations in a classical statistical mechanics model:

\begin{equation}\label{appendix:nega_x}
E_N = \log Z_A -  \log \tilde{Z}_A
\end{equation} 

\section{Calculation of SPT order parameters}\label{append:SPT}
Since the negativity spectrum is given by the wave functions of an SPT-cluster state under certain perturbations, it is natural to explore the connection between the persistence of long-range entanglement and the robustness of the SPTs. Here we show that these two are indeed intimately connected. 

\subsection{Pauli-Z decoherence}
In this case, the negativity spectrum (Eq.\ref{appendix:z_noise_spectrum}) is given by the wave functions in $X$ product-state basis of the perturbed state $\ket{ \tilde{\psi}   } \propto e^{ \beta_z \sum_{ i \in R_A} \prod_{j \in \partial^* i }Z_{B,j} } \ket{\psi}$, with $\ket{\psi} = U_{\text{CZ}} \ket{+}$ and $U_{\text{CZ}} =  \prod_{\expval{ i \in R_A, j \in R_B }   } \text{CZ}_{i, j}$. The potential singularity of $\ket{\tilde{\psi }}$ may be detected by computing the overlap $\bra{\tilde{\psi }} \ket{\tilde{\psi}}  \propto\bra{+}  U_{\text{CZ}}^{\dagger}   e^{ 2\beta_z \sum_{ i \in R_A} \prod_{j \in \partial^* i }Z_{B,j} }  U_{\text{CZ}}  \ket{+}=\bra{+}    e^{ 2\beta_z \sum_{ i \in R_A} \prod_{j \in \partial^* i }Z_{B,j} }  \ket{+} \propto \sum_{\tau}  e^{ 2\beta_z \sum_{ i \in R_A} \prod_{j \in \partial^* i }   \tau_j }$, which is the partition function of the statistical mechanics model that characterizes the entanglement negativity of the decohered mixed state (Eq.\ref{append:negativity_z}), albeit at different inverse temperatures. This provides a direct connection between the disentangling transition of the decohered state and the stability of the emergent SPTs in the fictitious Hilbert space.

The stability of the emergent SPTs can also be diagnosed by physical observables. For the 1d perturbed boundary state $\ket{\tilde{ \psi}  } \propto  e^{\beta_z \sum_{i} Z_{B,i}Z_{B,i+1}}\ket{\psi}$ that appears in the negativity spectrum of the 2d toric code, the $\mathbb{Z}_2\times \mathbb{Z}_2$ SPT order manifests in two types of string order parameters: $S_A(i,j)= Z_{B,i}  X_{ A,i+1}\cdots X_{A,j} Z_{B,j}$ and $S_B(i,j) = Z_{A,i}  X_{B,i} \cdots X_{B,j-1} Z_{A,j}$. For $S_A(i,j)$ operators, since it commutes with the perturbation $e^{\beta_z \sum_i Z_{B,i }Z_{B,i+1}   }$, one finds $\bra{\tilde{\psi}}  S_{A}(i,j) \ket{ \tilde{\psi} } =1 $, which is insensitive to the perturbation. On the other hand, $ \expval{S_B(i,j)}$ will be perturbed; since $S_B(i,j) e^{\beta_z \sum_n Z_{B,n }Z_{B,n+1}   } =e^{\beta_z \sum_n J_n Z_{B,n }Z_{B,n+1}   } S_B(i,j)$ where $J_n = -1 $ for $n=i-1, j-1$, and $J_n =1$ otherwise, one finds

\begin{equation}
\expval{S_B(i,j)}  = \frac{ \bra{\psi}e^{2\beta_z \sum_{n \neq  i-1, j-1} Z_{B,n }Z_{B,n+1}   }    \ket{\psi}      }{ \bra{\psi}e^{2\beta_z \sum_n Z_{B,n }Z_{B,n+1}   } \ket{\psi}} 
\end{equation}
Using the condition $\ket{\psi}  = U_{\text{CZ}}  \ket{+}$ with $ U_{\text{CZ}}$ being the depth-1 circuits of neighboring controlled-Z gates that creates the 1d cluster state, one has

\begin{equation}
\begin{split}
\expval{S_B(i,j)}  &= \frac{ \bra{+}e^{2\beta_z \sum_{n \neq  i-1, j-1} Z_{B,n }Z_{B,n+1}   }    \ket{+}      }{ \bra{+}e^{2\beta_z \sum_n Z_{B,n }Z_{B,n+1}   } \ket{+}} \\
&  =\frac{ \sum_{ \tau } e^{2\beta_z \sum_{n \neq  i-1, j-1} \tau_n \tau_{n+1}  }    }{ \sum_{ \tau } e^{2\beta_z \sum_{n} \tau_n \tau_{n+1}  }},  
\end{split}
\end{equation}
which is the ratio of the two partition functions of the 1d Ising model, with and without the Ising couplings at two endpoints of a string. Due to the absence of the finite-temperature transition of the 1d Ising model, this quantity is non-singular at any finite temperature, indicating the persistence of the SPT order in the state $e^{\beta_z \sum_{i }Z_{B,i}Z_{B,i+1}   } \ket{\psi}$ for any non-infinite $\beta_z$. Indeed, it is straightforward to evaluate the partition functions and find $ \expval{S_B(i,j)} = \frac{1}{ \left[\cosh(2\beta_z) \right]^2 \left[  1+ \tanh(2\beta_z)  \right]^L } >0 $ for any finite $\beta_z$.

\subsection{Pauli-X decoherence}
In this case, the negativity spectrum (Eq.\ref{append:x_spectrum}) is given by the wave functions in $X$ product-state basis of the perturbed state $\ket{ \tilde{\psi}   } \propto e^{ \beta_x \sum_{ j  \in R_B} \prod_{i  \in \partial j  }Z_{A,i} } \ket{\psi}$, with $\ket{\psi} = U_{\text{CZ}} \ket{+}$ and $U_{\text{CZ}} =  \prod_{\expval{ i \in R_A, j \in R_B }   } \text{CZ}_{i, j}$. Again, the potential singularity of $\ket{\tilde{\psi }}$ may be detected by computing the overlap $\bra{\tilde{\psi }} \ket{\tilde{\psi}}  \propto \bra{+}  U_{\text{CZ}}^{\dagger}   e^{ 2\beta_x \sum_{ j  \in R_B} \prod_{i \in \partial  j }Z_{A,i} }  U_{\text{CZ}}  \ket{+}=\bra{+}    e^{ 2\beta_x \sum_{ j  \in R_B} \prod_{i \in \partial j }Z_{A,i} }  \ket{+} \propto\sum_{\sigma}  e^{ 2\beta_x \sum_{j\in R_B} \prod_{i \in \partial j }  \sigma_i}$, which is the partition function of the statistical mechanics model that describes the negativity of the decohered mixed state (Eq.\ref{appendix:nega_x}), albeit at different inverse temperatures. This provides a direct connection between the disentangling transition of the decohered state and the stability of the emergent SPTs in the fictitious Hilbert space.

\section{Quantum-classcial mapping}\label{append:qc_mapping}
As discussed in the main text, the topological entanglement negativity of the 4d toric code subject to boundary decoherence relates to the subleading term of the free energy of the 3d classical Ising gauge theory at finite temperature, which is $2\log 2 $ at the low-temperature (deconfined) phase and $0$ at the high-temperature (confined) phase. Here, using a standard quantum-classical mapping (see e.g. \cite{1979_kogut,hsieh_2012_mapping}), we derive this subleading term by mapping the 3d classical Ising gauge theory at finite temperature to the 2d quantum Ising gauge theory at zero temperature.

First, we start from the 2d quantum Ising gauge theory described by the Hamiltonian $H= - \sum_p \prod_{e\in \partial p} Z_e - \Gamma \sum_e X_e$, where the first term denotes the product of Pauli-Zs on the four edges around a plaquette $p$, and the second term is the onsite transverse field. We compute $\tr \left[e^{\tilde{\beta} H } G \right]  $, where $G=\prod_v \delta_v(XXXX=1)$ enforces the Gauss-law constraint that the product of Pauli-Xs on the four edges emanating from any vertex $v$ is fixed at 1. To proceed, one can follow a standard Trotterization by inserting a complete Pauli-Z basis: 

\begin{equation} 
\begin{split}
\tr \left[e^{\tilde{\beta} H } G \right] &
= \lim_{M \to\infty}  \tr\left[ G
e^{ \frac{\tilde{\beta}\Gamma   }{M}  \sum_e X_e }   e^{\frac{\tilde{\beta}}{M} \sum_p \prod_{e\in \partial p} Z_e }  \right]^{M}  \\
&= \lim_{M \to\infty}  \sum_{ \{ s_{e,\tau}^{z} \}   }  \prod_{\tau=1}^M \bra{ \{ s^z_{e,\tau +1}  \} } Ge^{ \frac{\tilde{\beta}\Gamma   }{M}  \sum _e X_e   }e^{\frac{\tilde{\beta}}{M} \sum_p \prod_{e\in \partial p} Z_e } \ket{\{ s^z_{e,\tau}\}}
\end{split}
\end{equation} 
where $\tau$ labels the coordinate in the imaginary time direction, with the periodic boundary conditions imposed: $s_{e,M+1}^z   \equiv s_{e,1}^z   $. For each  matrix element, one has

\begin{equation} 
\begin{split} \bra{ \{ s^z_{e,\tau +1}  \} }G   e^{ \frac{\tilde{\beta}\Gamma   }{M}  \sum _e X_e   }  e^{\frac{\tilde{\beta}}{M} \sum_p \prod_{e\in \partial p} Z_e }\ket{\{ s^z_{e,\tau}\}}   &=  \bra{ \{ s^z_{e,\tau +1}  \} }G   e^{ \frac{\tilde{\beta}\Gamma   }{M}  \sum _e X_e   } \ket{\{ s^z_{e,\tau}\}} e^{\frac{\tilde{\beta}}{M} \sum_p \prod_{e\in \partial p} s^z_{e,\tau} }. 
\end{split}
\end{equation} 
By inserting a complete Pauli-X basis, the above quantity can be expressed as

\begin{equation} 
\begin{split}
&\sum_{ \{s_{e,\tau}^x \}}   \bra{ \{ s^z_{e,\tau +1}  \} } \ket{\{ s^x_{e,\tau}  \} }   \bra{ \{ s^x_{e,\tau}  \} } \ket{\{ s^z_{e,\tau}  \} }  
e^{ \frac{\tilde{\beta}\Gamma   }{M}  \sum _e s^x_{e,\tau}   } e^{\frac{\tilde{\beta}}{M}    \sum_p \prod_{e\in \partial p} s^z_{e,\tau} }\prod_v \frac{1+  \prod_{e \in \partial^* v} s^x_{e,\tau}}{2}   \\
&= 2^{-2L^2} \sum_{ \{s_{e,\tau}^x \}}  \prod_e \left( s_{e,\tau}^x\right)^{\frac{1-s_{e,\tau}^z}{2} +  \frac{1-s_{e,\tau+1}^z}{2}} 
e^{ \frac{\tilde{\beta}\Gamma   }{M}  \sum _e s^x_{e,\tau}   } e^{\frac{\tilde{\beta}}{M}    \sum_p \prod_{e\in \partial p} s^z_{e,\tau} }  2^{-L^2} \sum_{\{\lambda_{v,\tau+\frac{1}{2}}  \}}  \prod_v \left( \prod_{e \in \partial^*  v}    s_{e,\tau}^x \right)^{\frac{1-\lambda_{v, \tau+ \frac{1}{2} }  }{2}}.
\end{split}
\end{equation} 
Note that we introduce $\lambda_{v,\tau+\frac{1}{2}} = \pm 1$ as a Lagrange multiplier to resolve the Gauss law constraint, and these variables will become the spins on the edge along the imaginary time direction, in between the imaginary time $\tau$ and $\tau+1$.

Now one can sum over $s_{e,\tau}^x$; for a given $s_{e,\tau}^{x}$, one computes the summation: 

\begin{equation}
\begin{split}
&\sum_{ s_{e,\tau}^x  }  \left( s_{e,\tau}^x\right)^{\frac{1-s_{e,\tau}^z}{2} +  \frac{1-s_{e,\tau+1}^z}{2} + \frac{1-\lambda_{v,\tau+\frac{1}{2}}}{2} + \frac{1-\lambda_{v',\tau+\frac{1}{2}}}{2} 
  } e^{  \frac{ \tilde{\beta}\Gamma  }{M} s_{e,\tau}^x } \\
  &=   e^{  \frac{ \tilde{\beta}\Gamma  }{M}} + s_{e,\tau}^z  s^z_{e,\tau+1}  \lambda_{v,\tau+\frac{1}{2}} \lambda_{v',\tau+\frac{1}{2}}   e^{  - \frac{ \tilde{\beta}\Gamma }{M}} \\
  &= e^{ \beta_\tau s_{e,\tau}^z  s^z_{e,\tau+1}  \lambda_{v,\tau+\frac{1}{2}} \lambda_{v',\tau+\frac{1}{2}}   }
\end{split}
\end{equation}
with $v, v'$ being two vertices on the boundary of the edge $e$, and $\beta_\tau \equiv -\frac{1}{2}  \log [\tanh( \frac{\tilde{\beta} \Gamma  }{M}  )  ] $. This term can be understood as the four-spin coupling around the plaquette spanned by one space direction and one temporal direction. Using the results above, finally, we find 

\begin{equation}
\tr\left[ e^{\tilde{\beta} H} G\right] = 
\lim_{M \to \infty}  2^{-3L^2M}  \sum_{\sigma_e}  e^{\sum_p K_p  \sigma \sigma\sigma \sigma   },  
\end{equation}
where $K_p = \frac{ \tilde{\beta}  }{M}$ for plaquettes spanned by two spatial directions, and $K_p =\beta_\tau  = -\frac{1}{2}  \log [\tanh( \frac{\tilde{\beta} \Gamma  }{M}  )  ]$ for the plaquettes spanned by one spatial direction and one temporal direction.  

Taking the limit $\tilde{\beta} \to \infty$ with $\tilde{\beta}/M = \beta = O(1) $ fixed, the L.H.S. reduces to the ground-subspace projector, and the R.H.S. becomes the classical 3d Ising Gauge theory at finite temperature $\beta$: 

\begin{equation}
N_g  e^{- \tilde{\beta}E_g}  = 2^{ -3L^2 \tilde{\beta} \beta }   \sum_{ \sigma_e}e^{ \beta \sum_p \sigma \sigma  \sigma  \sigma   }, 
\end{equation}
where $N_g, E_g$ are the ground-state degeneracy and energy of the 2d quantum gauge theory, and we have also considered the isotropic limit by setting $\beta =  -\frac{1}{2} \log[  \tanh(\beta \Gamma) ]  $. This completes the mapping between the 3d classical Ising gauge theory at finite temperature and the 2d quantum Ising gauge theory at zero temperature. In particular, tuning the inverse temperature $\beta$ in the 3d classical model amounts to tuning $\Gamma$ in the 2d quantum model, which exhibits a confinement-deconfinement transition at a critical $\Gamma_c$, with $N_g=4$ ($\Gamma<\Gamma_c$), and $N_g=1$ ($\Gamma>\Gamma_c$). This implies the free energy of the 3d classical model takes the form $\log Z= L^3 f + \log g$, with $g=4$ in the low-temperature phase and $g=1$ in the high-temperature phase.

\section{Toric code subject to both X,Z  decoherence}\label{append:xz_both}

Here we consider the 2d toric code under both Pauli-Z and Pauli-X boundary decoherence with noise rate $p_z$ and $p_x$, and discuss the negativity spectrum. 

In the 2d toric code, along a 1d boundary of size $L$, there are $L$ star and $L$ plaquette operators, which we label in order: $A_1, B_1, A_2, B_2 , \cdots$. The $Z$ decoherence occurs on the edge shared between two adjacent $B_i, B_{i+1}$ plaquette stabilizers, and the $X$ decoherence occurs on the vertices shared between two adjacent $A_i, A_{i+1}$ star stabilizers. Correspondingly, the boundary part of the decohered mixed state reads  

\begin{equation}
\rho_\partial  \propto \sum_{a,b} \prod_i A_i^{a_i}\prod_i B_i^{b_i}   e^{\beta_x  \sum_{i=1}^L \sigma_i  \sigma_{i+1}   }e^{\beta_z  \sum_{i=1}^L \tau_i  \tau_{i+1}   }
\end{equation}
with $\sigma_i = (-1)^{a_i}\in\{ 1,-1 \}$ and $\tau_i = (-1)^{b_i} \in\{ 1,-1 \}$. Therefore, the negativity spectrum can be expressed as

\begin{equation}
\begin{split}
\rho^{\Gamma}  \propto \bra{+} \prod_i Z_{A,i}^{\frac{1-A_i}{2}}  \prod_i Z_{B,i}^{\frac{1-B_i}{2}}    e^{\beta_x \sum_{i} Z_{A,i}Z_{A,i+1 }  } e^{\beta_z \sum_{i} Z_{B,i}Z_{B,i+1 }  } \ket{\psi}, 
\end{split}
\end{equation}
where $\ket{\psi}$ is the 1d cluster state defined in the fictitious Hilbert space.

This may be simplified by using the condition $ Z_{B,i}Z_{B,i+1 } \ket{\psi} = X_{A,i+1} \ket{\psi} $ and  $ Z_{A,i}Z_{A,i+1 } \ket{\psi} = X_{B,i} \ket{\psi}$: 

\begin{equation}
\begin{split}
e^{\beta_x \sum_{i} Z_{A,i}Z_{A,i+1 }  } e^{\beta_z \sum_{i} Z_{B,i}Z_{B,i+1 }  } \ket{\psi} 
&=  e^{\beta_x \sum_{i} Z_{A,i}Z_{A,i+1 }  } e^{\beta_z \sum_{i} X_{A,i}  } \ket{\psi} \\ 
&=  (\cosh\beta_x)^L \sum_{\tau} \prod_i ( \tanh\beta_x Z_{A,i} Z_{A,i+1}  
 )^{ \frac{1-\tau_i}{2}  } e^{\beta_z \sum_i X_{A,i}} \ket{\psi} \\
 &  =(\cosh\beta_x)^L \sum_{\tau} e^{\beta_z \sum_i \tau_{i-1}\tau_i X_{A,i}} \prod_i ( \tanh\beta_x X_{B,i} )^{ \frac{1-\tau_i}{2}  } \ket{\psi}, 
\end{split}
\end{equation}
with $\tau \equiv  \{\tau_i =\pm 1 \}$. As a result, we obtain the final expression of the negativity spectrum:

\begin{equation}
\rho^{\Gamma} = \frac{1}{Z}\chi(A,B)\sum_\tau e^{\beta_z \sum_i \tau_{i-1}\tau_i A_i + K_x\sum_i  \tau_i   }\prod_i \tau_i^{\frac{1-B_i}{2}  }, 
\end{equation}
where $K_x = -\frac{1}{2} \log (\tanh \beta_x) $, $Z$ is a normalization constant to ensure that negativity spectrum sums to one, and $\chi(A,B) = \frac{\bra{+} 
 \prod_i Z_{A,i }^{\frac{1-A_i}{2}}   \prod_i Z_{B,i }^{\frac{1-B_i}{2}}   \ket{\psi}     }{ \bra{+} \ket{\psi}    }$ is the strange correlator (or equivalently, the ratio of wave functions in X basis) of the 1d cluster state. This expression indicates that the negativity spectrum is given by $\chi(A,B)$ multiplied by the $\tau$ spin correlations in a 1d Ising model under a symmetry-breaking field, where $A_i, B_i$ determine the sign of the Ising coupling and spin insertion, respectively. In particular, the sign structure of the negativity spectrum is no longer solely determined by $\chi(A,B)$, which poses a technical difficulty in deriving the entanglement negativity directly from the negativity spectrum.

\end{document}